# Wavelength Diversity Expands the Expressive Power of Optical Computing

Shiqi Chen[1,2,3], Che-Yung Shen[1,2,3], Parnian Ghapandar Kashani[1,2,3], Yongkang Cheng[1,2,3], Aydogan Ozcan[1,2,3*]

[1] Electrical and Computer Engineering Department, University of California, Los Angeles, CA, 90095, USA

[2] Bioengineering Department, University of California, Los Angeles, CA, 90095, USA

[3] California NanoSystems Institute (CNSI), University of California, Los Angeles, CA, 90095, USA

* ozcan@ucla.edu

## Abstract

Universal function approximation underpins neural networks' capacity to represent nonlinear relationships, but realizing this capability in optical hardware requires engineering the computational features encoded in light. Here, we show that wavelength diversity enhances optical function-approximation expressivity through wavelength-scaled Fourier features generated by a shared phase encoder, without increasing the number of input-encoded harmonics. Multiple wavelengths contribute to the representation of each output function rather than merely serving as independent computational channels. We establish theoretical scaling laws and approximation-error bounds, revealing that, for nondispersive encoders, the spectral endpoints determine the accessible computational bandwidth, while intermediate illumination wavelengths enrich the available harmonics within that bandwidth. Numerical studies demonstrate improved approximation of real- and complex-valued nonlinear functions, with a representative design achieving more than a 100-fold reduction in mean-squared error relative to tested monochromatic optical processors with the same number of trainable phase features. Experimentally, we validate enhanced approximation accuracy and demonstrate the simultaneous approximation of 4,136 distinct nonlinear functions using a compact, single-layer diffractive processor under five-wavelength illumination. These results establish wavelength diversity as an additional physical resource for scaling the expressive power of optical computing and enhancing the accuracy of parallel nonlinear function approximation.

## Introduction

Nonlinear function approximation is central to scientific computing, signal processing and machine learning. Universal approximation theory[1–7] establishes that neural networks with sufficient capacity can approximate broad classes of target functions, while approximation scaling laws quantify how the resources required to achieve a prescribed accuracy grow with target-function complexity. A complementary body of work on Fourier-feature mappings and periodic neural representations has demonstrated the importance of computational feature selection for accurately representing complex functions and signals[8–10]. Translating these mathematical capabilities into physical computing systems[11–25] requires understanding not only the available hardware degrees of freedom but also the diversity of computational features that can be physically represented. Diffractive optical processors provide an emerging platform for parallel nonlinear function approximation through phase encoding and passive optical propagation[26–28]. In this framework, the input wavefront supplies computational Fourier features, while optimized diffractive structures synthesize their coefficients through spatially varying complex-valued point-spread functions (PSFs)[29–32]. The expressive power of these optical processors is therefore governed by both the computational features encoded in light and the available optical degrees of freedom, linking attainable approximation accuracy to the physical resources of the processor[27,33].

For a monochromatic diffractive function approximator, the prescribed input phase encoding determines the computational Fourier-frequency set, while square-law detection generates harmonics corresponding to pairwise differences between these frequencies. Increasing the number of trainable phase features can improve the synthesis of the associated spatially varying complex PSF coefficients, but does not by itself expand the available detector-level frequency support imposed by the input encoding and readout scheme. Approximation accuracy can therefore remain limited when target nonlinear functions require frequency components that are inadequately represented by the available computational Fourier features. This limitation highlights a fundamental distinction between increasing the number of trainable parameters and expanding the computational feature space, motivating alternative approaches to scaling the expressive power of optical computing.

Illumination-wavelength diversity provides such an approach. Because the phase delay produced by a physical encoder depends on wavelength, simultaneously illuminating the same encoder at multiple wavelengths generates scaled computational Fourier features without increasing the number of input-encoded harmonics. The diffractive processor weights these features through its wavelength-dependent, spatially varying coherent response, and the wavelength-specific intensity contributions are summed at each output detector. For a nondispersive encoder, the spectral endpoints determine the accessible computational bandwidth, while intermediate wavelengths enrich the set of distinct detector harmonics within that bandwidth. Wavelength diversity can therefore expand computational bandwidth and improve nonlinear function-representation accuracy, provided that the required wavelength-dependent coherent responses can be synthesized with sufficient accuracy. Importantly, these computational Fourier harmonics are defined over the input-variable space and should not be confused with the spatial-frequency components of propagating optical fields.

Here, we theoretically and experimentally demonstrate how wavelength diversity expands the expressive power of optical processors. Specifically, we develop spectral-domain diffractive processors with expanded computational bandwidth and enhanced nonlinear function-representation capacity. We establish a theoretical framework that describes the generation of wavelength-scaled computational Fourier features, incorporates encoder dispersion and wavelength-dependent detector response, and derives approximation-error bounds accounting for truncation and PSF synthesis errors. Numerical studies demonstrate substantially improved approximation of non-negative real-valued and complex-valued nonlinear functions, including comparisons with monochromatic processors having larger numbers of trainable phase features. We further extend this framework to compact, single-layer encoder–decoder co-localized diffractive processors and experimentally demonstrate improved nonlinear function-approximation accuracy by introducing additional illumination wavelengths within fixed spectral endpoints. Using simultaneous five-wavelength illumination spanning 500–700 nm, we experimentally demonstrate the parallel approximation of 4,136 nonlinear functions. In situ learning further reduces the experimental approximation error and narrows the error distribution across the output channels, bringing the measured performance closer to numerical predictions. Together, these theoretical and experimental findings establish wavelength diversity as an additional physical resource for expanding the computational capabilities and expressive power of optical computing systems.

## Results

**Figure 1** illustrates the operating principle of a spectral-domain diffractive optical processor, in which simultaneous illumination at multiple wavelengths enables a shared phase encoder to generate wavelength-scaled computational Fourier features. An optimized diffractive processor weights these features through its wavelength-dependent, spatially varying coherent PSFs, and the resulting intensity contributions are combined at each output detector to synthesize nonlinear functions. This wavelength-dependent expansion of computational Fourier features provides an additional physical resource for increasing the expressive power of diffractive optical processors. As detailed below, the resulting improvements in nonlinear function-representation capacity depend on the available computational bandwidth, the diversity of accessible harmonics, and the accuracy with which the required wavelength-dependent PSFs can be synthesized.

Importantly, the presented approach is fundamentally different from wavelength multiplexing schemes that increase the number of nonlinear functions implemented in parallel by assigning different functions to different wavelength channels[26,34]. Such wavelength multiplexing approaches increase computational parallelism without inherently enhancing the computational bandwidth, approximation accuracy, or nonlinear function representation capacity of each individual output function, whereas our approach exploits wavelength diversity to expand the available computational Fourier features and improve the approximation accuracy of each output function.

Furthermore, these spectral-domain diffractive processors synthesize nonlinear functions using computational Fourier harmonics defined over abstract input variables, distinct from the spatial-frequency components of propagating optical fields in conventional Fourier optics. This distinction allows the framework to extend to nonlinear functions of more than three independent variables, with computational dimensionality set by the encoded input space rather than the three-dimensional geometry of light propagation.

## Optical forward model of spectral-domain diffractive processors

Instead of a monochromatic diffractive function approximator, here we consider $N_w$ wavelengths ($\lambda_i$, $i = 1,2, \dots, N_w$) *simultaneously* illuminating the same encoder plane and being summed in intensity at an output detector $k$; see **Fig. 1**. The central conclusion of our analysis is that the same physical encoder generates a wavelength-scaled family of computational frequencies. If the wavelength-dependent spatially varying PSF matrices can be synthesized with sufficiently small error, which requires increasing the number ($N$) of trainable phase features of the diffractive processor in proportion to $N_w$, the increased detector-level computational bandwidth enhances the nonlinear function representation capability of the spectral-domain diffractive optical processor, as we theoretically analyze, quantify, and experimentally demonstrate in this section.

The symbol $\lambda_i$ denotes the physical illumination wavelength. The symbol $\Lambda$ refers to the computational Fourier-frequency set in the abstract input-variable space $\mathbf{a} \in \mathbb{R}^{\mathrm{D}}$. These computational frequencies are not the transverse spatial frequencies of the propagating optical field[27]. The physical wavelengths are ordered as $\lambda_1 < \lambda_2 < \cdots < \lambda_{N_w}$, where $\lambda_1$ is the shortest wavelength and $\lambda_{N_w}$ is the longest wavelength. The compact domain of the nonlinear-function argument is denoted by $\mathcal{K} \subset \mathbb{R}^D$. The key symbols used in this work are summarized in Table 1.

| | |
|---|---|
| $\mathbf{a} = (a_1, \dots, a_D) \in \mathbb{R}^D$ | Input argument of the target nonlinear function(s). |
| $p = 1, \dots, N_p$ | Phase encoder-channel index. |
| $k = 1, \dots, N_f$ | Output channel number or target-function index. |
| $i = 1, \dots, N_w$ | Illumination-wavelength index. $\lambda_{N_w}$ is the longest wavelength. |
| $\lambda_i$ | Wavelength of illumination channel $i$. |
| $n(\lambda_i)$ | Wavelength-dependent phase-delay factor of the encoder at $\lambda_i$. |
| $\boldsymbol{\alpha}_p \in \mathbb{R}^D$ | Computational Fourier-frequency vector assigned to encoder channel $p$ at the reference wavelength, which is taken as the longest wavelength, $\lambda_{N_w}$. |
| $\Lambda = \{\boldsymbol{\alpha}_p\}_{p=1}^{N_p}$ | Reference computational frequency set at the longest wavelength, $\lambda_{N_w}$. |
| $K_i$ | Wavelength-dependent scaling of the computational frequencies. |
| $\hat{F}(k; p, \lambda_i)$ | Wavelength-dependent spatially varying coherent PSFs at the illumination wavelength $\lambda_i$, connecting the input channel $p$ to the output channel $k$. |
| $\mu_{i,k}$ | Wavelength-dependent scaling constant at detector $k$ for illumination wavelength $\lambda_i$. |
| $N$ | Total number of trainable phase-only diffractive features used in the diffractive processor. |
| $N_w$ | Total number of illumination wavelengths that are simultaneously on. |
| $N_f$ | Total number of nonlinear functions implemented in parallel by the diffractive processor. |
| $N_p$ | Total number of effective/distinct encoded Fourier-feature channels at the input. |

**Table 1**: A summary of the key symbols and definitions used in this manuscript for spectral-domain diffractive function approximators. The bandwidth in our formulas refers to a computational Fourier bandwidth in the abstract function-variable space $\mathbf{a}$. It should not be confused with the physical transverse spatial-frequency bandwidth of propagating electromagnetic fields.

At one illumination wavelength, the coherent Fourier-feature encoder and output-field model can be expressed as[27]:

$$u_{\text{in}}(p; \mathbf{a}) = \exp\left(j2\pi\boldsymbol{\alpha}_p \cdot \mathbf{a}\right), \text{and } \tilde{f}_k(\mathbf{a}) = \sum_{p=1}^{N_p} \hat{F}\,(k; p)\exp\left(j2\pi\boldsymbol{\alpha}_p \cdot \mathbf{a}\right) \quad \textbf{(1A)}$$

Let $h_p(\mathbf{a})$ denote the physical encoder modulation associated with input channel $p$. It may represent, e.g., a piston-based micro-mirror array, a spatial light modulator (SLM) pixel, an electrically controlled optical-path length, or another calibrated quantity that creates a dynamic phase delay per pixel. At the illumination wavelength $\lambda_i$, we assume that the encoder phase is:

$$\phi_{p,i}(\mathbf{a}) = \frac{2\pi}{\lambda_i} n(\lambda_i) h_p(\mathbf{a}) \quad \textbf{(1B)}$$

Here, $n(\lambda_i)$ denotes a wavelength-dependent effective phase-delay factor, which is an encoder architecture-dependent factor. For example, for a reflective piston-based micro-mirror array (phase modulator) located in a dielectric medium with a given dispersion, the round-trip path-length per encoder pixel is understood to be absorbed into $h_p(\mathbf{a})$, and therefore $n(\lambda_i)$ indicates the refractive index of the dielectric medium at $\lambda_i$.

We further assume that the phase encoder device is designed so that the longest wavelength $\lambda_{N_w}$ implements the reference Fourier feature used in our analysis:

$$\phi_{p,N_w}(\mathbf{a}) = 2\pi \boldsymbol{\alpha}_p \cdot \mathbf{a}. \quad \textbf{(2A)}$$

Equating the two descriptions at $\lambda_{N_w}$ gives $\frac{2\pi}{\lambda_{N_w}} n(\lambda_{N_w}) h_p(\mathbf{a}) = 2\pi \boldsymbol{\alpha}_p \cdot \mathbf{a}$, and therefore we can get:

$$\phi_{p,i}(\mathbf{a}) = \frac{2\pi}{\lambda_i} n(\lambda_i) \frac{\lambda_{N_W}}{n(\lambda_{N_W})} \boldsymbol{\alpha}_p \cdot \mathbf{a} = 2\pi \left[\frac{\lambda_{N_W}}{\lambda_i} \frac{n(\lambda_i)}{n(\lambda_{N_W})}\right] \boldsymbol{\alpha}_p \cdot \mathbf{a} \quad \textbf{(2B)}$$

Based on this, we can define the wavelength-dependent computational-frequency scaling factor ($K_i$) as:

$$K_i \equiv \frac{\lambda_{N_W}}{\lambda_i} \frac{n(\lambda_i)}{n(\lambda_{N_W})} \quad \textbf{(2C)}$$

By definition, we have $K_{N_w} = 1$. The complex field encoded by the input channel $p$ at wavelength $\lambda_i$ is therefore given by:

$$u_{\text{in}}(p; \mathbf{a}, \lambda_i) = \exp\left(j2\pi K_i \boldsymbol{\alpha}_p \cdot \mathbf{a}\right);\ i = 1,2, \ldots, N_w \quad \textbf{(2D)}$$

Based on Eq. (2D), one can see that, for a spectral-domain diffractive processor, the same physical phase encoder pattern produces $N_w$ scaled computational Fourier features at the $N_w$ unique illumination wavelengths that are simultaneously on.

## Wavelength-scaled computational frequencies of a spectral-domain diffractive processor

The reference computational frequency set at $\lambda_{N_w}$ is given as: $\Lambda = \{\boldsymbol{\alpha}_p\}_{p=1}^{N_p}$. Based on Eq. (2D), at the illumination wavelength $\lambda_i$ that is simultaneously on, the same phase encoder generates:

$$\Lambda_i = K_i \Lambda = \{K_i \boldsymbol{\alpha}_p\}_{p=1}^{N_p} \quad \textbf{(3A)}$$

The aggregate field-level frequency set supplied by the simultaneous wavelength illumination is given by:

$$\Lambda_{\text{field}}^{(N_w)} = \bigcup_{i=1}^{N_w} K_i \Lambda \quad \textbf{(3B)}$$

If we define the maximum norm of a reference field frequency as $\alpha_{\lambda_{N_w}}^{\max} = \max_{\boldsymbol{\alpha} \in \Lambda} \| \boldsymbol{\alpha} \|$, then the maximum field frequency at wavelength $\lambda_i$ is $\alpha_{\lambda_i}^{\max} = K_i \alpha_{\lambda_{N_w}}^{\max}$. Consequently, the maximum field-level computational frequency across the entire set of active illumination wavelength channels is given by:

$$\alpha_{\text{field}}^{\max} = K_{\max} \alpha_{\lambda_{N_w}}^{\max}, \quad K_{\max} = \max_{1 \le i \le N_w} K_i \quad \textbf{(3C)}$$

The number of distinct computational-frequency vectors obeys $\left|\Lambda_{\text{field}}^{(N_w)}\right| \le N_w N_p$, with equality (i.e., $\left|\Lambda_{\text{field}}^{(N_w)}\right| = N_w N_p$) only satisfied when the scaled frequency sets do not overlap.

These results show that wavelength diversity expands the available computational Fourier-feature set without increasing the number of input-encoded harmonics. Quite importantly, *adding illumination wavelengths between fixed spectral endpoints can increase the density and diversity of the spectral feature set for nonlinear function approximation even when it does not change the outermost frequency*; this will be further detailed and discussed in our analyses below.

## Wavelength-dependent spatially varying coherent PSFs and output spectral field synthesis

The forward operator of a spectral-domain diffractive network (**Fig. 1**) depends on wavelength through both optical propagation and wavelength-dependent diffractive phase modulation. Let $\hat{F}(k; p, \lambda_i)$ denote the effective spatially varying coherent PSF coefficient connecting encoder channel $p$ to output channel $k$ at wavelength $\lambda_i$. The complex field at output channel $k$ can be written as:

$$\tilde{f}_{k,i}(\mathbf{a}) = \sum_{p=1}^{N_p} \hat{F}(k; p, \lambda_i) \exp(j 2\pi K_i \boldsymbol{\alpha}_p \cdot \mathbf{a}) \quad \textbf{(4A)}$$

In Eq. (4A), the vector $K_i \boldsymbol{\alpha}_p$ is the wavelength-scaled computational frequency, whereas $\hat{F}(k; p, \lambda_i)$ represents the spatially varying complex coefficients (i.e., the PSFs) synthesized by the optimized spectral-domain diffractive processor, which is frozen/fixed after its training/optimization. The larger computational frequencies are created by $K_i$, while the wavelength dependence of $\hat{F}(k; p, \lambda_i)$ permits different complex weights to be assigned to those scaled features.

For a desired wavelength-specific field, we can define the finite Fourier *target/ground truth* nonlinear function as:

$$f_{k,i,\Lambda_i}(\mathbf{a}) = \sum_{p=1}^{N_p} F_{k,i}\left(K_i \boldsymbol{\alpha}_p\right) \exp\left(j2\pi K_i \boldsymbol{\alpha}_p \cdot \mathbf{a}\right) \quad \textbf{(4B)}$$

Therefore, an accurate field-based function approximation requires $\hat{F}(k;p,\lambda_i) \approx F_{k,i}(K_i \boldsymbol{\alpha}_p)$ for all $i$, $p$, and $k$. We can define the PSF synthesis error as:

$$E_{\mathrm{PSF},k,i} = \sum_{p=1}^{N_p} \left| F_{k,i}(K_i \boldsymbol{\alpha}_p) - \hat{F}(k;p,\lambda_i) \right|, \quad \textbf{(4C)}$$

which yields[27]:

$$\left\| f_{k,i,\Lambda_i} - \tilde{f}_{k,i} \right\|_{\infty,\mathcal{K}} \leq E_{\mathrm{PSF},k,i} \quad \textbf{(4D)}$$

For the illumination wavelength $\lambda_i$, we can define the input feature vector as:

$$\mathbf{v}_i(\mathbf{a}) = \begin{bmatrix} \exp(j2\pi K_i \boldsymbol{\alpha}_1 \cdot \mathbf{a}) \\ \exp(j2\pi K_i \boldsymbol{\alpha}_2 \cdot \mathbf{a}) \\ \vdots \\ \exp\left(j2\pi K_i \boldsymbol{\alpha}_{N_p} \cdot \mathbf{a}\right) \end{bmatrix} \in \mathbb{C}^{N_p},$$

and the wavelength-dependent coefficient matrix as: $\hat{\mathbf{F}}_i = \left[\hat{F}(k;p,\lambda_i)\right] \in \mathbb{C}^{N_f \times N_p}$.

The $N_f$ output fields at the output of a spectral-domain diffractive optical function approximator at the illumination wavelength $\lambda_i$ can be written as:

$$\tilde{\mathbf{f}}_i(\mathbf{a}) = \hat{\mathbf{F}}_i \mathbf{v}_i(\mathbf{a}) \quad \textbf{(5A)}$$

Since we consider a passive, time-invariant spectral-domain diffractive optical processor, the wavelength-labeled forward map is block diagonal:

$$\begin{bmatrix} \tilde{\mathbf{f}}_1 \\ \tilde{\mathbf{f}}_2 \\ \vdots \\ \tilde{\mathbf{f}}_{N_w} \end{bmatrix} = \begin{bmatrix} \hat{\mathbf{F}}_1 & \mathbf{0} & \cdots & \mathbf{0} \\ \mathbf{0} & \hat{\mathbf{F}}_2 & \cdots & \mathbf{0} \\ \vdots & \vdots & \ddots & \vdots \\ \mathbf{0} & \mathbf{0} & \cdots & \hat{\mathbf{F}}_{N_w} \end{bmatrix} \begin{bmatrix} \mathbf{v}_1 \\ \mathbf{v}_2 \\ \vdots \\ \mathbf{v}_{N_w} \end{bmatrix} \quad \textbf{(5B)}$$

The nonzero blocks contain $N_w N_p N_f$ complex-valued coefficient entries. Our earlier works on wavelength-multiplexed[31,35,36] phase-only diffractive optical processors demonstrated that we need:

$$N \gtrsim 2 N_w N_p N_f \quad \textbf{(5C)}$$

to achieve $\hat{F}(k;p,\lambda_i) \approx F_{k,i}(K_i \boldsymbol{\alpha}_p)$ for all $i = 1,\ldots,N_w$ , $p = 1,\ldots,N_p$ and $k = 1,\ldots,N_f$. This is a parameter-count and design heuristic, and the realized PSF error (Eqs. (4C-D)) depends on, for example, the diffractive architecture/geometry, number of diffractive layers, aperture, wavelength span, optimization landscape, material dispersion, fabrication tolerances, throughput, and numerical aperture. Therefore, while wavelength diversity expands the computational features available for nonlinear function approximation, realizing the corresponding gains in expressive power requires sufficient trainable optical degrees of freedom to accurately synthesize their wavelength-dependent spatially varying PSF coefficients.

## Nonlinear function approximation via output intensity of a spectral-domain diffractive processor

We assume that, for each channel $k$, the field $\tilde{f}_{k,i}(\mathbf{a})$ at the output of the spectral-domain diffractive processor is designed to approximate $f_{k,i,\Lambda_i}(\mathbf{a})$ for $i = 1, \ldots, N_w$ (see Eq. (4D)), and that the photodetector assigned to channel $k$ performs a time-averaged square-law detection, where distinct optical wavelengths that are simultaneously on add in intensity. Let $\mu_{i,k} > 0$ be the wavelength-dependent scaling constant at detector $k$ (which includes in its definition, e.g., the detector response and diffraction efficiency at wavelength $\lambda_i$), we can then write the detected output signal function as:

$$\tilde{g}_k^{(N_w)}(\mathbf{a}) = \sum_{i=1}^{N_w} \mu_{i,k} \left|\tilde{f}_{k,i}(\mathbf{a})\right|^2 \quad \textbf{(6A)}$$

Using Equation (4A), $\tilde{f}_{k,i}(\mathbf{a}) = \sum_{p=1}^{N_p} \hat{F}\,(k; p, \lambda_i)\exp\left(j2\pi K_i \boldsymbol{\alpha}_p \cdot \mathbf{a}\right)$, we can get:

$$\begin{aligned}\tilde{g}_k^{(N_w)}(\mathbf{a}) &= \sum_{i=1}^{N_w} \sum_{p,q=1}^{N_p} \mu_{i,k}\, \hat{F}(k; p, \lambda_i)\hat{F}(k; q, \lambda_i)^* \exp\left[j2\pi K_i\left(\boldsymbol{\alpha}_p - \boldsymbol{\alpha}_q\right) \cdot \mathbf{a}\right] \\ &= \sum_{i=1}^{N_w} \mu_{i,k} \left|\sum_{p=1}^{N_p} \hat{F}(k; p, \lambda_i)\exp\left[j2\pi K_i \boldsymbol{\alpha}_p \cdot \mathbf{a}\right]\right|^2\end{aligned} \quad \textbf{(6B)}$$

This equation shows that square-law detection creates computational harmonics of the form $K_i\left(\boldsymbol{\alpha}_p - \boldsymbol{\alpha}_q\right)$. If we define the reference difference set as: $\Lambda - \Lambda = \left\{\boldsymbol{\alpha}_p - \boldsymbol{\alpha}_q : \boldsymbol{\alpha}_p, \boldsymbol{\alpha}_q \in \Lambda\right\}$, then the **available detector-level frequency support** is defined as:

$$\Lambda_{\text{det}}^{(N_w)} \equiv \bigcup_{i=1}^{N_w} K_i(\Lambda - \Lambda) \quad \textbf{(6C)}$$

If $\mathbf{0} \in \Lambda$, then $K_i\Lambda \subseteq K_i(\Lambda - \Lambda)$ since $K_i\boldsymbol{\alpha}_p = K_i\left(\boldsymbol{\alpha}_p - \mathbf{0}\right)$. Thus, interference at the output detector with a nonzero DC component causes the field-level-scaled harmonics to appear directly in the detected intensity. If $\Lambda$ is symmetric and contains both $\boldsymbol{\alpha}$ and $-\boldsymbol{\alpha}$, the intensity detection at the output can also generate $K_i[\boldsymbol{\alpha} - (-\boldsymbol{\alpha})] = 2K_i\boldsymbol{\alpha}$. That is why a symmetric field-frequency set can give a detector bandwidth twice the maximum field-frequency radius.

The resulting detector-level computational-frequency set, $\Lambda_{\text{det}}^{(N_w)}$, determines the harmonics available for nonlinear function approximation. Wavelength diversity can expand this set by extending its frequency support and enriching the available harmonics within a fixed bandwidth, thereby increasing the range of nonlinear functions that the processor can accurately represent.

Equation (6B) can be written as a finite Fourier expansion:

$$\tilde{g}_k^{(N_w)}(\mathbf{a}) = \sum_{\boldsymbol{\beta} \in \Lambda_{\text{det}}^{(N_w)}} \hat{G}_k\,(\boldsymbol{\beta})\exp(j2\pi\boldsymbol{\beta} \cdot \mathbf{a}) = \sum_{i=1}^{N_w} \mu_{i,k} \left|\tilde{f}_{k,i}(\mathbf{a})\right|^2 \quad \textbf{(6D)}$$

where $K_i\left(\boldsymbol{\alpha}_p - \boldsymbol{\alpha}_q\right) = \boldsymbol{\beta}$ and

$$\hat{G}_k(\boldsymbol{\beta}) = \sum_{\substack{i,p,q: \\ K_i(\boldsymbol{\alpha}_p - \boldsymbol{\alpha}_q) = \boldsymbol{\beta}}} \mu_{i,k}\, \hat{F}(k; p, \lambda_i)\hat{F}(k; q, \lambda_i)^*$$

$$\sum_{\boldsymbol{\beta}\in\Lambda_{\mathrm{det}}^{(N_w)}} \hat{G}_k(\boldsymbol{\beta}) = \sum_{i=1}^{N_w}\sum_{p,q=1}^{N_p} \mu_{i,k}\,\hat{F}(k;p,\lambda_i)\hat{F}(k;q,\lambda_i)^* = \tilde{g}_k^{(N_w)}(\mathbf{0})$$

$$= \sum_{i=1}^{N_w} \mu_{i,k} \left|\sum_{p=1}^{N_p} \hat{F}(k;p,\lambda_i)\right|^2 \quad \textbf{(6E)}$$

If we consider a desired/target non-negative finite Fourier function that can be expressed through the wavelength-specific target fields defined in Eq. (4B), we then get:

$$g_{k,\Lambda_{\mathrm{det}}}(\mathbf{a}) = \sum_{\boldsymbol{\beta}\in\Lambda_{\mathrm{det}}^{(N_w)}} G_k(\boldsymbol{\beta})\exp(j2\pi\boldsymbol{\beta}\cdot\mathbf{a}) = \sum_{i=1}^{N_w}\mu_{i,k}\left|f_{k,i,\Lambda_i}(\mathbf{a})\right|^2, \quad \textbf{(6F)}$$

where

$$G_k(\boldsymbol{\beta}) = \sum_{\substack{i,p,q:\\ K_i(\boldsymbol{\alpha}_p-\boldsymbol{\alpha}_q)=\boldsymbol{\beta}}} \mu_{i,k}\, F_{k,i}(K_i\boldsymbol{\alpha}_p)F_{k,i}(K_i\boldsymbol{\alpha}_q)^*$$

This target nonlinear function can be accurately approximated through the forward model of the spectral-domain diffractive processor if $\hat{G}_k(\boldsymbol{\beta}) \approx G_k(\boldsymbol{\beta})$ for all $\boldsymbol{\beta}$. As a sufficiency constraint, the spectral-domain diffractive processor can approximate this representation when $\hat{F}(k;p,\lambda_i) \approx F_{k,i}(K_i\boldsymbol{\alpha}_p)$ for all $i = 1,\dots,N_w$, $p = 1,\dots,N_p$ and $k = 1,\dots,N_f$ by using $N \gtrsim 2N_wN_pN_f$ degrees of freedom; see Eq. (5C). To support this conclusion, **Fig. 2** reports the function approximation accuracy of spectral-domain diffractive processors for $N_f = 25$, $N_p = 17$, and $\alpha_{\lambda_{N_w}}^{\max} = \max_{\boldsymbol{\alpha}\in\Lambda} \|\boldsymbol{\alpha}\| = 8$, with mean squared error (MSE) values compared as a function $N_w$ and $N$ as well as $\alpha_{\mathrm{G}}^{\max}$ which refers to the target (ground truth) functions' maximum frequency content. A similar analysis supporting the same conclusions is also reported in **Supplementary Fig. S1** for increased values of $\alpha_{\mathrm{G}}^{\max}$ along with an increased number of encoder harmonics, i.e., $N_p = 33$ and $\alpha_{\lambda_{N_w}}^{\max} = 16$. Both of these figures confirm that adding illumination wavelengths between fixed spectral endpoints (400 nm and 600 nm in these examples) increases the density and diversity of the spectral feature set for substantially improving the nonlinear function approximation accuracy.

Our analyses also reveal that these conclusions on the function representation advantages of illumination wavelength diversity in a spectral-domain diffractive processor hold for not only the case of $N = 2N_wN_pN_f$ but also for $N = N_wN_pN_f$ and $N = 3N_wN_pN_f$. In addition to the analyses reported in **Fig. 2** and **Supplementary Fig. S1**, this spectral-domain function representation improvement can also be visualized in **Fig. 3** and **Supplementary Figs. S2-6**, which compare some examples of the output of the spectral-domain diffractive processors against the ground truth target functions, showing significant improvements as $N_w$ increases.

Earlier results shown in **Figs. 2-3** and **Supplementary Figs. S1-S6** demonstrate that direct intensity detection yields an accurate approximation of non-negative real functions. Signed or complex-valued nonlinear target functions require, for example, a differential or multi-intensity readout convention[27]. To showcase this capability and extend our intensity-based readout to complex-valued nonlinear function approximation, we used three spatially separated output detectors, each assigned to a non-negative real-valued component function and a fixed complex basis coefficient, $e_c = \exp(j2\pi c/3)$, where $c = 0,1,2$. Therefore, $N_d = 3N_f$, where $N_d$ refers to the number of output detector channels. Following Eq. (6A), each detector collects the summed intensity contributions from all simultaneously illuminated wavelengths: $\tilde{g}_{k,c}^{(N_w)}(\mathbf{a}) = \sum_{i=1}^{N_w} \mu_{i,k,c} \mid \tilde{f}_{k,c,i}(\mathbf{a}) \mid^2$. The complex-valued function estimate is reconstructed by combining these three detectors' real-valued signals: $\hat{h}_k^{(N_w)}(\mathbf{a}) = \sum_{c=0}^{2} e_c \, \tilde{g}_{k,c}^{(N_w)}(\mathbf{a})$. Because the three complex basis coefficients are separated by 120°, their non-negative linear combinations span the entire complex plane. This allows a complex-valued target function to be represented through three non-negative component functions without background subtraction from the individual detector signals. Each detector signal is synthesized using the wavelength-scaled computational features described above, extending the spectral-domain representation strategy to complex-valued nonlinear targets; see **Figs. 4-5** and **Supplementary Figs. S7-S8** for numerical results that demonstrate significant improvements in the complex function approximation accuracy of spectral-domain diffractive processors as $N_w$ increases.

These numerical results demonstrate that wavelength diversity can enhance nonlinear function-representation accuracy beyond what is achievable by simply increasing the number of trainable phase features in the tested monochromatic reference processors. Moreover, the accuracy improvements obtained by adding intermediate wavelengths within fixed spectral endpoints highlight the importance of harmonic diversity, beyond the expansion of computational bandwidth alone.

## Computational bandwidth expansion and harmonic enrichment through wavelength diversity

In relation to Eqs. (6B-C), let's define the diameter of the reference computational frequency set for monochromatic illumination at $\lambda_{N_w}$ as

$$B_{\Lambda} = \max_{\boldsymbol{\alpha}_p, \boldsymbol{\alpha}_q \in \Lambda} \|\boldsymbol{\alpha}_p - \boldsymbol{\alpha}_q\| \quad \textbf{(7A)}$$

Stated differently, this term, $B_{\Lambda}$, refers to the largest computational frequency norm that can be generated through square-law detection at the reference illumination wavelength $\lambda_{N_w}$.

Based on Eqs. (6B-C), the *available detector-level computational bandwidth* is given by:

$$B_{\text{det}}^{(N_w)} = K_{\max} B_{\Lambda} \quad \textbf{(7B)}$$

Since $\boldsymbol{\beta} = K_i(\boldsymbol{\alpha}_p - \boldsymbol{\alpha}_q)$, we can also write $\| \boldsymbol{\beta} \| \leq K_{\max} B_{\Lambda}$. If the reference frequency set is symmetric, $\Lambda = -\Lambda$, then we have $B_{\Lambda} = 2\alpha_{\lambda_{N_w}}^{\max}$, which yields:

$$B_{\text{det}}^{(N_w)} = 2K_{\max}\alpha_{\lambda_{N_w}}^{\max} \quad \textbf{(7C)}$$

Based on these equations, the frequency endpoint-scaling factor ($S_{\text{BW}}$) relative to the longest wavelength of the multi-wavelength illumination can be given as:

$$S_{\mathrm{BW}} \equiv \frac{B_{\mathrm{det}}^{(N_w)}}{B_\Lambda} = K_{\max} = \max_{1 \le i \le N_w} K_i \quad \textbf{(7D)}$$

Using Equation (2C), for a general dispersive phase encoder device, we get:

$$S_{\mathrm{BW}} = \max_{1 \le i \le N_w} \left[ \frac{\lambda_{N_w}}{\lambda_i} \frac{n(\lambda_i)}{n(\lambda_{N_w})} \right] \quad \textbf{(7E)}$$

Equations (7D-E) indicate that if the wavelength-dependent spatially varying PSFs can be synthesized by the optimized spectral-domain diffractive processor with sufficiently small error (i.e., $\hat{F}(k; p, \lambda_i) \approx F_{k,i}(K_i \boldsymbol{\alpha}_p)$ for all $i$, $p$ and $k$, see Eq. (4D)), the available detector-level computational bandwidth increases by $K_{\max}$.

If the shortest wavelength supplies the largest value of $n(\lambda_i)/\lambda_i$, then Eqs. (7D-E) can be simplified as:

$$S_{\mathrm{BW}} = \frac{\lambda_{N_w}}{\lambda_1} \frac{n(\lambda_1)}{n(\lambda_{N_w})} \quad \textbf{(7F)}$$

In the general dispersive phase encoder case, the wavelength ratio alone is not sufficient to determine $S_{\mathrm{BW}}$. If dispersion of the phase encoder device is nonmonotonic, the maximizing wavelength need not be the shortest one, and Eqs. (7D-E) should be used. Equation (7F) can be further simplified if we assume a dispersion-free phase encoder device, such as a piston-based micro-mirror array in air, also known as a Phase Light Modulator (PLM)[37]. In such a dispersion-free phase encoder, we get: $S_{\mathrm{BW}} = K_{\max} = \frac{\lambda_{N_w}}{\lambda_1}$.

Importantly, we should emphasize that adding intermediate wavelengths, not just the shortest wavelength, is important because it supplies more separately weighted spectral components, increasing the number of distinct detector harmonics; see Eq. (6C), $\Lambda_{\mathrm{det}}^{(N_w)} \equiv \bigcup_{i=1}^{N_w} K_i(\Lambda - \Lambda)$. Stated differently, adding intermediate wavelengths can increase the number and density of distinct detector harmonics and can provide additional wavelength-specific coefficient contributions. Therefore, in the nondispersive case, or more generally when $K_i$ varies monotonically over the illumination wavelengths, the spectral endpoints determine the outer radius, while the intermediate wavelengths determine how richly that radius is populated and weighted. In the general dispersive case, the outer radius is determined by the active illumination wavelength that maximizes $K_i$.

These theoretical results identify two complementary mechanisms through which wavelength diversity expands the expressive power of diffractive optical processors: extending the accessible computational-frequency support and enriching the available harmonic representation within that support. Both mechanisms contribute to the improved nonlinear function-approximation accuracy observed in our numerical studies (**Figs. 2–5** and **Supplementary Figs. S1–S8**).

## Approximation-error bound of spectral-domain diffractive processors

Following Eq. (6F), the desired non-negative detector function is given by:

$$g_{k,\Lambda_{\text{det}}}(\mathbf{a}) = \sum_{\boldsymbol{\beta}} G_k(\boldsymbol{\beta})\exp(j2\pi\boldsymbol{\beta}\cdot\mathbf{a}) = \sum_{i=1}^{N_w} \mu_{i,k} \left|f_{k,i,\Lambda_i}(\mathbf{a})\right|^2$$

We can define the wavelength-specific complex-field approximation error as[27]:

$\epsilon_{k,i} = \|f_{k,i,\Lambda_i} - \tilde{f}_{k,i}\|_{\infty,\mathcal{K}}$, where $\mathcal{K} \subset \mathbb{R}^D$ represents the compact domain of the nonlinear-function argument, $\mathbf{a} = (a_1, \dots, a_D)$.

For each wavelength, we have:

$$\left||f_{k,i,\Lambda_i}|^2 - |\tilde{f}_{k,i}|^2\right| \leq |f_{k,i,\Lambda_i} - \tilde{f}_{k,i}|\left(|f_{k,i,\Lambda_i}| + |\tilde{f}_{k,i}|\right).$$

Using the definition of $\epsilon_{k,i}$ gives:

$$\left||f_{k,i,\Lambda_i}|^2 - |\tilde{f}_{k,i}|^2\right| \leq 2\epsilon_{k,i}\|f_{k,i,\Lambda_i}\|_{\infty,\mathcal{K}} + \epsilon_{k,i}^2.$$

Summing the weighted errors over all the illumination wavelengths that are simultaneously on gives:

$$\left\|g_{k,\Lambda_{\text{det}}} - \tilde{g}_k^{(N_w)}\right\|_{\infty,\mathcal{K}} \leq \sum_{i=1}^{N_w} \mu_{i,k}\left(2\epsilon_{k,i}\|f_{k,i,\Lambda_i}\|_{\infty,\mathcal{K}} + \epsilon_{k,i}^2\right) \quad \textbf{(8A)}$$

For a general non-negative target nonlinear function $g_k$ that is not exactly in this finite function class, we can also define the truncation[27] error as:

$E_{\text{trunc},k}^{(N_w)} = \|g_k - g_{k,\Lambda_{\text{det}}}\|_{\infty,\mathcal{K}}$, which yields:

$$\left\|g_k - \tilde{g}_k^{(N_w)}\right\|_{\infty,\mathcal{K}} \leq E_{\text{trunc},k}^{(N_w)} + \sum_{i=1}^{N_w} \mu_{i,k}\left(2\epsilon_{k,i}\|f_{k,i,\Lambda_i}\|_{\infty,\mathcal{K}} + \epsilon_{k,i}^2\right) \quad \textbf{(8B)}$$

A more complete physical error budget can be defined by using additional noise terms (due to, e.g., illumination phase error or other factors that are implementation dependent).

These error bounds distinguish the limitations imposed by the available computational Fourier representation from those arising from imperfect synthesis of the required wavelength-dependent spatially varying coherent responses. They, therefore, provide a theoretical basis for understanding how wavelength diversity and trainable optical degrees of freedom jointly determine achievable nonlinear function-approximation accuracy.

## Expanding the expressive power of single-layer spectral-domain diffractive optical processors

The preceding analysis was formulated for a general spectral-domain diffractive function approximator whose output is described through spatially varying coherent PSF coefficients at different illumination wavelengths, synthesized through optimized and cascaded multi-layer phase structures (see **Fig. 1**). We next demonstrate that the computational bandwidth advantages of multi-wavelength illumination described in this work also extend to single-layer Encoder-Decoder (E+D) co-localized diffractive function approximators[28]. The encoder-decoder co-localized (E+D) processor architecture places a dynamic, *input-* *dependent* encoder (E) and an optimized *static* (i.e., *input-independent*) decoder (D) on the same phase-only diffractive plane; see **Fig. 6**. After a free-space propagation step, the encoder and decoder fields interfere at the output detector, and square-law detection converts their coherent mixing term into a programmable real-valued nonlinear function.

To shed more light on the multi-wavelength illumination advantage in a single-layer spectral-domain E+D diffractive function approximator architecture, let's assume that the physical detector window assigned to output channel $k$ be $\mathcal{W}_k$, with width $w_k$. The detected output signal of an E+D spectral-domain diffractive architecture can then be written as[28]:

$$\tilde{g}_k^{(N_w)}(\mathbf{a}) = \sum_{i=1}^{N_w} \mu_{i,k} \int_{\mathcal{W}_k} \left|E_{\mathrm{enc},i}(u;\mathbf{a}) + E_{\mathrm{dec},i}(u)\right|^2 \mathrm{d}u$$

$$= \sum_{i=1}^{N_w} \mu_{i,k} \left[\int_{\mathcal{W}_k} \left|E_{\mathrm{enc},i}(u;\mathbf{a})\right|^2 \mathrm{d}u + \int_{\mathcal{W}_k} \left|E_{\mathrm{dec},i}(u)\right|^2 \mathrm{d}u + 2\, Re\left\{\int_{\mathcal{W}_k} E_{\mathrm{enc},i}(u;\mathbf{a}) E_{\mathrm{dec},i}(u)^* \,\mathrm{d}u\right\}\right]$$

**(9A)**

Here, $E_{\mathrm{enc},i}(u;\mathbf{a})$ and $E_{\mathrm{dec},i}(u)$ indicate the encoder-generated and decoder-generated output fields, respectively, at an illumination wavelength of $\lambda_i$. If we assume $h_i$ denotes the *free-space* coherent PSF between the E+D phase encoder plane and the output plane of the optical processor at wavelength $\lambda_i$, we can then write:

$$E_{\mathrm{enc},i}(u;\mathbf{a}) = \sum_{p\in E} h_i(u;p) \exp\left(j2\pi K_i \boldsymbol{\alpha}_p \cdot \mathbf{a} + jK_i\, \mathrm{b}_p\right)$$

$$E_{\mathrm{dec},i}(u) = \sum_{p'\in D} h_i(u;p') \exp\left(jK_i \theta_{p'}\right) \quad \textbf{(9B)}$$

where $\theta_{p'}$ and $\mathrm{b}_p$ are trainable phase terms at the illumination wavelength $\lambda_{N_w}$, and are both independent of the input $\mathbf{a}$; they represent the optimized decoder ($D$) phase distribution and the phase bias term within the encoder ($E$) region, respectively. Because the encoder and decoder are implemented on the same co-localized phase modulator plane, the same physical phase-scaling factor $K_i$ applies to both regions as indicated in Eq. (9B). In this notation, $E$ and $D$ regions do not spatially overlap; the phase bias term at $\lambda_i$, i.e., $K_i\, \mathrm{b}_p$, physically serves as the decoder within the $E$ region. This way, any arbitrary topology of encoder and decoder regions is physically covered in our forward model described here. Note that, unlike the spatially varying PSFs ($\hat{F}(k;p,\lambda_i)$) corresponding to the multi-layer diffractive processors (**Fig. 1**) described in earlier sections, $h_i$ represents a shift-invariant complex-valued PSF due to the free-space propagation.

Equations (9A-B) reflect the $N_w$-wavelength extension of the monochromatic E+D diffractive processor architecture. For every illumination wavelength, it contains the encoder self-intensity term ($\int_{\mathcal{W}_k} \left|E_{\mathrm{enc},i}(u;\mathbf{a})\right|^2 \mathrm{d}u$), the decoder-only background ($\int_{\mathcal{W}_k} \left|E_{\mathrm{dec},i}(u)\right|^2 \mathrm{d}u$), and the programmable encoder-decoder mixing term ($2\, Re\left\{\int_{\mathcal{W}_k} E_{\mathrm{enc},i}(u;\mathbf{a}) E_{\mathrm{dec},i}(u)^* \,\mathrm{d}u\right\}$). This latter encoder–decoder mixing term provides directly programmable contributions at the encoded computational frequencies, while the encoder self-intensity term contributes additional difference-frequency components to the full detector response toward the nonlinear function-synthesis capability of an E+D architecture. Specifically, this mixing term in Eq. (9A) can be expanded as:

$$M_{k,i}(\mathbf{a}) = 2\,Re\left\{\int_{\mathcal{W}_k} E_{\mathrm{enc},i}(u;\mathbf{a})E_{\mathrm{dec},i}(u)^*\,\mathrm{d}u\right\}$$

$$= 2Re\left\{\int_{\mathcal{W}_k}\sum_{p\in E} h_i\,(u;p)\exp\left(j2\pi K_i\boldsymbol{\alpha}_p\cdot\mathbf{a}+jK_i\,\mathrm{b}_p\right)E_{\mathrm{dec},i}(u)^*\,du\right\}$$

$$= 2Re\left\{\sum_{p\in\mathrm{E}} C_{k,i,p}\exp\left(j2\pi K_i\boldsymbol{\alpha}_p\cdot\mathbf{a}\right)\right\} \quad \textbf{(9C)}$$

where $C_{k,i,p} \equiv \int_{\mathcal{W}_k} h_i\,(u;p)\exp\left(jK_i\,\mathrm{b}_p\right)E_{\mathrm{dec},i}(u)^*\,du$.

Based on Eq. (9C), the available spectral support of the mixing term at one illumination wavelength satisfies:

$$\mathrm{spec}_{\mathbf{a}}\left[M_{k,i}\right] \subseteq K_i\Lambda \cup (-K_i\Lambda) \quad \textbf{(9D)}$$

Taking the union over all active illumination wavelengths yields the following expression for the available frequency support of the mixing term:

$$\Lambda_{\mathrm{mix}}^{(N_w)} = \bigcup_{i=1}^{N_w}\left[K_i\Lambda \cup (-K_i\Lambda)\right] \quad \textbf{(9E)}$$

If $\Lambda = -\Lambda$, then Eq. (9E) simplifies to:

$$\Lambda_{\mathrm{mix}}^{(N_w)} = \bigcup_{i=1}^{N_w} K_i\Lambda \quad \textbf{(9F)}$$

Similarly, by taking the union over all active illumination wavelengths, we can express the available frequency support of the encoder self-intensity term ($\int_{\mathcal{W}_k}\left|E_{\mathrm{enc},i}(u;\mathbf{a})\right|^2\mathrm{d}u$) and the decoder-only background ($\int_{\mathcal{W}_k}\left|E_{\mathrm{dec},i}(u)\right|^2\mathrm{d}u$) term as follows:

$$\Lambda_{\mathrm{self}}^{(N_w)} = \bigcup_{i=1}^{N_w} K_i(\Lambda - \Lambda) \quad \textbf{(9G)}$$

$$\Lambda_{\mathrm{dec-only}}^{(N_w)} = \{\mathbf{0}\} \quad \textbf{(9H)}$$

The full detector signal, $\tilde{g}_k^{(N_w)}(\mathbf{a})$ in Eq. (9A), is the sum of the decoder-only background, mixing term, and self-intensity term. Therefore, the Fourier support of its sum is contained in the union of the individual supports, which yields the following expression for the available frequency support of $\tilde{g}_k^{(N_w)}(\mathbf{a})$ under the E+D spectral-domain diffractive processor architecture:

$$\Lambda_{\mathrm{E+D}}^{(N_w)} = \{\mathbf{0}\} \cup \Lambda_{\mathrm{mix}}^{(N_w)} \cup \Lambda_{\mathrm{self}}^{(N_w)} = \Lambda_{\mathrm{mix}}^{(N_w)} \cup \Lambda_{\mathrm{self}}^{(N_w)} \quad \textbf{(9I)}$$

Here and in our analysis below, $\Lambda_{\mathrm{E+D}}^{(N_w)}$, $\Lambda_{\mathrm{mix}}^{(N_w)}$, and $\Lambda_{\mathrm{self}}^{(N_w)}$ denote the architecture-level available frequency sets. Following our earlier notation, we have $\alpha_{\lambda_{N_w}}^{\max} \equiv \max_{\boldsymbol{\alpha} \in \Lambda} \parallel \boldsymbol{\alpha} \parallel$ and $B_\Lambda \equiv \max_{\boldsymbol{\alpha}_p, \boldsymbol{\alpha}_q \in \Lambda} \parallel \boldsymbol{\alpha}_p - \boldsymbol{\alpha}_q \parallel$ for the reference illumination wavelength, $\lambda_{N_w}$. Based on these equations and Eqs. (9E-I), we can get the following expressions for the available bandwidth across all illumination wavelengths for each term:

$$B_{\text{mix}}^{(N_w)} = K_{\max} B_{\text{mix}}^{(\lambda_{N_w})} = K_{\max} \alpha_{\lambda_{N_w}}^{\max} \text{ with } B_{\text{mix}}^{(\lambda_{N_w})} = \alpha_{\lambda_{N_w}}^{\max} \quad \textbf{(9J)}$$

$$B_{\text{self}}^{(N_w)} = K_{\max} B_{\text{self}}^{(\lambda_{N_w})} = K_{\max} B_\Lambda \text{ with } B_{\text{self}}^{(\lambda_{N_w})} = B_\Lambda \quad \textbf{(9K)}$$

$$B_{\text{E+D}}^{(N_w)} = K_{\max} B_{\text{E+D}}^{(\lambda_{N_w})} = K_{\max} \max\left\{\alpha_{\lambda_{N_w}}^{\max}, B_\Lambda\right\} \text{ with } B_{\text{E+D}}^{(\lambda_{N_w})} = \max\left\{\alpha_{\lambda_{N_w}}^{\max}, B_\Lambda\right\} \quad \textbf{(9L)}$$

Equations (9J-L) confirm that, same as in Eq. (7D), the frequency endpoint-scaling factor ($S_{\text{BW}}$) relative to the longest wavelength of the multi-wavelength illumination for the E+D processor is also given by:

$$S_{\text{BW}} = K_{\max} = \max_{1 \le i \le N_w} \left[\frac{\lambda_{N_w}}{\lambda_i} \frac{n(\lambda_i)}{n(\lambda_{N_w})}\right] \quad \textbf{(9M)}$$

To numerically test the predicted frequency scaling factor, i.e., $S_{\text{BW}} = K_{\max}$, and quantify its effect on the nonlinear function-approximation accuracy of a spectral-domain E+D diffractive processor, we conducted a comparison as shown in **Fig. 7** and **Supplementary Figs. S9-S11** (see **Methods** for details). Increasing the number of illumination wavelengths, while keeping the spectral endpoints at 500 and 700 nm, consistently improves approximation accuracy across the tested target nonlinear functions. This improvement persists when compared with monochromatic E+D reference architectures that have the same number of phase degrees of freedom, further supporting the enhanced function representation capacity of spectral-domain E+D diffractive processors. For example, at $\alpha_{\text{G}}^{\max} = 20$, the nine-wavelength E+D processor with $N = 2\, r\, N_w N_p N_f = 3 N_w N_p N_f$ ($r$=1.5) achieves an MSE of $\sim 3\times10^{-5}$, more than two orders of magnitude lower than the enlarged monochromatic reference E+D processors with the same number of trainable phase features (**Fig. 7a**). Furthermore, increasing $N_w$ from 2 to 9 within the same spectral endpoints reduces the MSE by ~40-fold in this case, demonstrating substantial function representation accuracy gains without further increasing the outer computational-frequency radius. The representative nonlinear function examples reported in **Supplementary Figs. S9–S11** further illustrate these improvements across different diffractive-feature budgets with $r$ = 0.5, 1.0, and 1.5, showing increasingly accurate reproduction of the target functions' oscillatory features, consistent with the enriched computational frequency coverage enabled by spectral diversity.

## Experimental Results

To experimentally validate the enhanced nonlinear function-representation capacity enabled by wavelength diversity, we implemented a compact, single-layer spectral-domain diffractive optical processor using a co-localized encoder–decoder architecture. We first investigated how increasing the number of illumination wavelengths within fixed spectral endpoints affects nonlinear function-approximation accuracy. **Figure 8a** shows the optical layout and experimental setup, in which a phase-only SLM implements the co-localized encoder and decoder (E+D), and a camera records the output intensities over spatially separated detector regions. To evaluate the effect of illumination-wavelength diversity experimentally, we initially compared diffractive processors approximating $N_f = 25$ target functions with $\alpha_G^{\max} = 20$, $N_p = 17$ and $\alpha_{\lambda_{N_w}}^{\max} = 8$ (**Fig. 8b**). The illumination configurations comprised monochromatic operation at 500 nm and 700 nm ($N_w = 1$ in each case), two-wavelength illumination at 500 nm and 700 nm ($N_w = 2$), and five-wavelength illumination spanning 500–700 nm in 50-nm increments ($N_w = 5$). **Figure 8b** compares the numerical designs, zero-shot experimental measurements, and measurements after in-situ learning (see the **Methods** for details). The five-wavelength diffractive processor ($N_w = 5$) achieved a zero-shot experimental MSE of $\sim 7.9 \times 10^{-3}$, which decreased to $\sim 4.9 \times 10^{-3}$ after in-situ learning, closely matching the numerical design. Following in-situ learning, its MSE was ~4-fold lower than that of the two-wavelength processor, with both configurations sharing the same spectral endpoints. Together with the numerical comparisons in **Fig. 7**, these measurements support the enriched computational frequency coverage enabled by spectral diversity, thereby enhancing the function representation capacity of diffractive processors. Furthermore, **Fig. 9** presents the experimentally measured function outputs after in-situ learning for all 25 target functions corresponding to the five-wavelength diffractive processor, showing a good agreement with the distinct oscillatory profiles of the target functions across the input interval.

Next, we demonstrated the parallel approximation of $N_f = 4{,}136$ target nonlinear functions with $\alpha_G^{\max} = 20$, $N_p = 17$ and $\alpha_{\lambda_{N_w}}^{\max} = 8$. The spatial assignment of the encoding harmonics, the optimized input-independent decoder phase pattern, and the detector layout are shown in **Supplementary Fig. S12**. **Figure 10a** compares the distributions of the function-wise MSE across all 4,136 output functions for the numerical design, its zero-shot experimental implementation, and the measurements after in-situ learning. In-situ learning reduced the median experimental MSE and narrowed the interquartile range, bringing the experimentally measured error distribution closer to the numerical predictions. **Figure 10b** further resolves the experimental performance spatially by reporting the mean MSE across detector subregions after in situ learning. The errors are lower in the central region and relatively higher near the detector-array boundary, revealing spatial variation in approximation accuracy across the output plane.

Furthermore, **Fig. 10c** presents 25 randomly selected target functions and their measured outputs, with additional examples provided in **Supplementary Figs. S13–S15**. These representative function responses complement the full-channel error statistics provided in **Fig. 10a** and the spatial error analysis in **Fig. 10b**. Together, these experiments demonstrate the enhanced nonlinear function-approximation accuracy enabled by wavelength diversity and its compatibility with massively parallel optical computation using a compact, single-layer spectral-domain diffractive processor.

## Discussion

Our findings reveal that the expressive power of diffractive optical computing can be expanded by engineering the diversity of computational features encoded in light, rather than relying solely on increasing the number of trainable optical degrees of freedom. By exploiting multiple illumination wavelengths, a shared phase encoder generates an expanded set of computational Fourier features without increasing the number of input-encoded harmonics. For nondispersive encoders, the spectral endpoints determine the maximum accessible computational bandwidth, while intermediate wavelengths can enrich the harmonic coverage and provide additional wavelength-dependent coefficient contributions within that bandwidth. These findings establish the illumination spectrum as an additional physical resource for controlling the nonlinear function-representation capacity of optical computing systems.

Importantly, the benefits of wavelength diversity depend on the ability of the diffractive processor to synthesize the required wavelength-dependent spatially varying coherent responses with sufficient accuracy. Our theoretical framework captures this dependence through approximation-error bounds and physical resource-scaling considerations. In practical implementations, factors such as encoder dispersion, finite phase-modulation range, optical throughput, detector response and experimental noise can further influence the achievable accuracy. Illumination wavelength diversity, therefore, provides an additional approach to scaling optical computational expressivity, complementing the available trainable hardware degrees of freedom rather than eliminating their importance.

Our experimental demonstrations using a compact, single-layer encoder–decoder architecture confirm the advantages of spectral diversity for nonlinear function approximation and establish its compatibility with massively parallel optical computation. While the present experiments focus on real-valued nonlinear functions of a single input variable, our theoretical framework extends to higher-dimensional input spaces, and our numerical studies also demonstrate corresponding benefits for complex-valued nonlinear functions. Future investigations could explore multivariable nonlinear mappings, functions learned from experimental data, and integrated photonic implementations that exploit wavelength diversity. More broadly, these findings introduce spectral diversity as a general design consideration for expanding the computational capabilities of light-based information-processing systems.

## Supplementary Information includes:

- **Methods**
    - Optical forward model of spectral-domain diffractive processors
    - Multi-wavelength illumination-based nonlinear function approximation
    - Training loss functions and implementation details

- Hyperparameter and geometry optimization of spectral-domain diffractive processors
- Details of the experimental setup

- **Supplementary Figures S1-S15**

## Figures

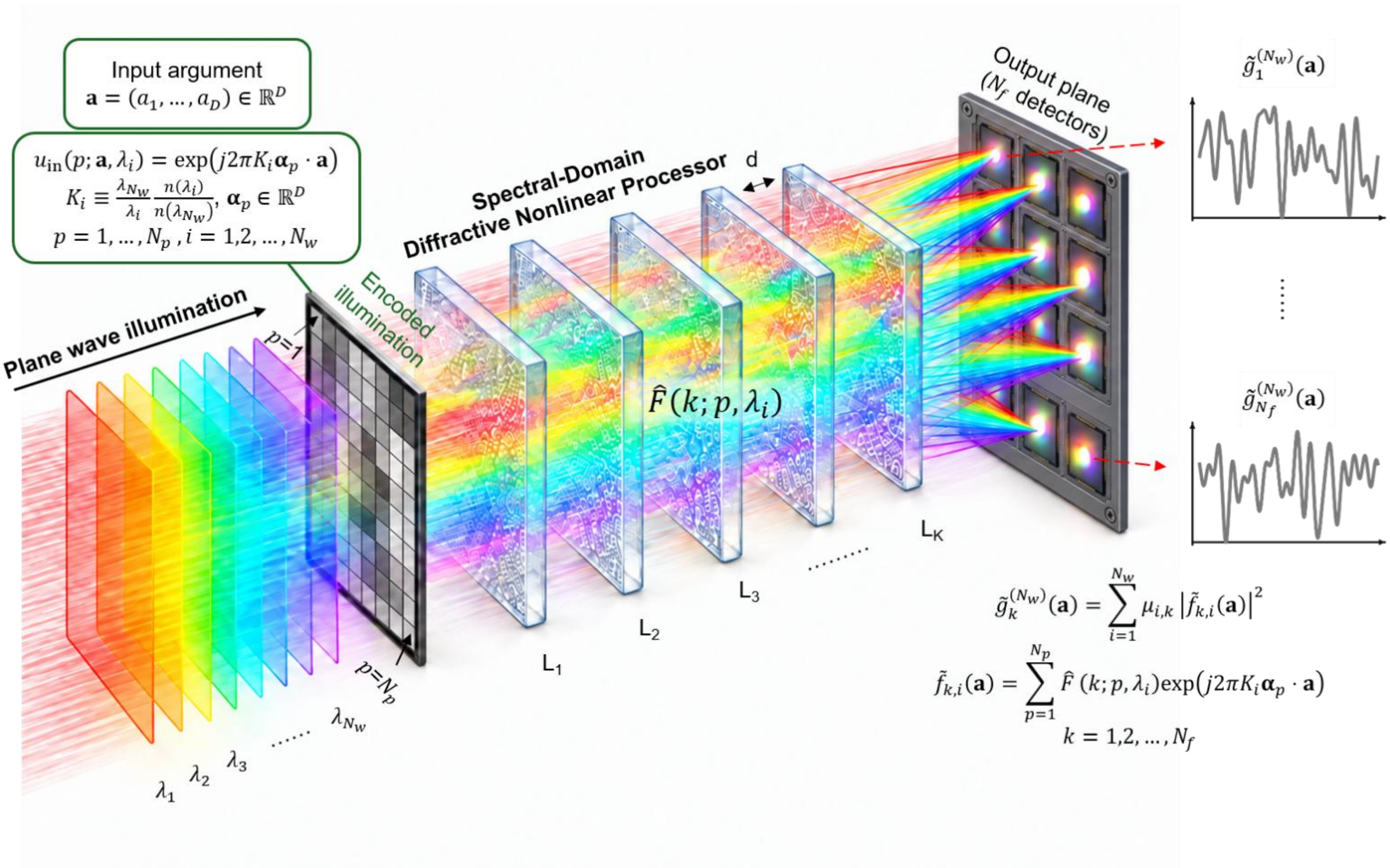

**Fig. 1 Spectral-domain diffractive nonlinear processor.** The input argument $\mathbf{a} = (a_1, \dots, a_D) \in \mathbb{R}^D$ is encoded by $N_p$ phase-only encoder channels and simultaneously illuminated at $N_w$ wavelengths. The resulting multi-spectral optical field $u_{\text{in}}(p; \mathbf{a}, \lambda_i)$ propagates through an optimized diffractive processor containing $N$ trainable phase-only features. The processor implements wavelength-dependent, spatially varying coherent PSFs, $\hat{F}(k; p, \lambda_i)$, which map each input channel $p$ to output channel $k$ at the illumination wavelength $\lambda_i$. The resulting fields are jointly transformed onto $N_f$ output detectors, each corresponding to a distinct target nonlinear function. At the output channel $k$, the detected intensities from distinct wavelengths are summed up with wavelength- and detector-dependent scaling factors $\mu_{i,k}$, yielding the nonlinear function output $\tilde{g}_k^{(N_w)}(\mathbf{a}) = \sum_{i=1}^{N_w} \mu_{i,k} \left|\tilde{f}_{k,i}(\boldsymbol{a})\right|^2$. $L_K$ refers to the number of diffractive layers within the spectral-domain diffractive processor. An initial version of this illustration was generated using ChatGPT with the authors' text prompt and was subsequently modified by the authors.

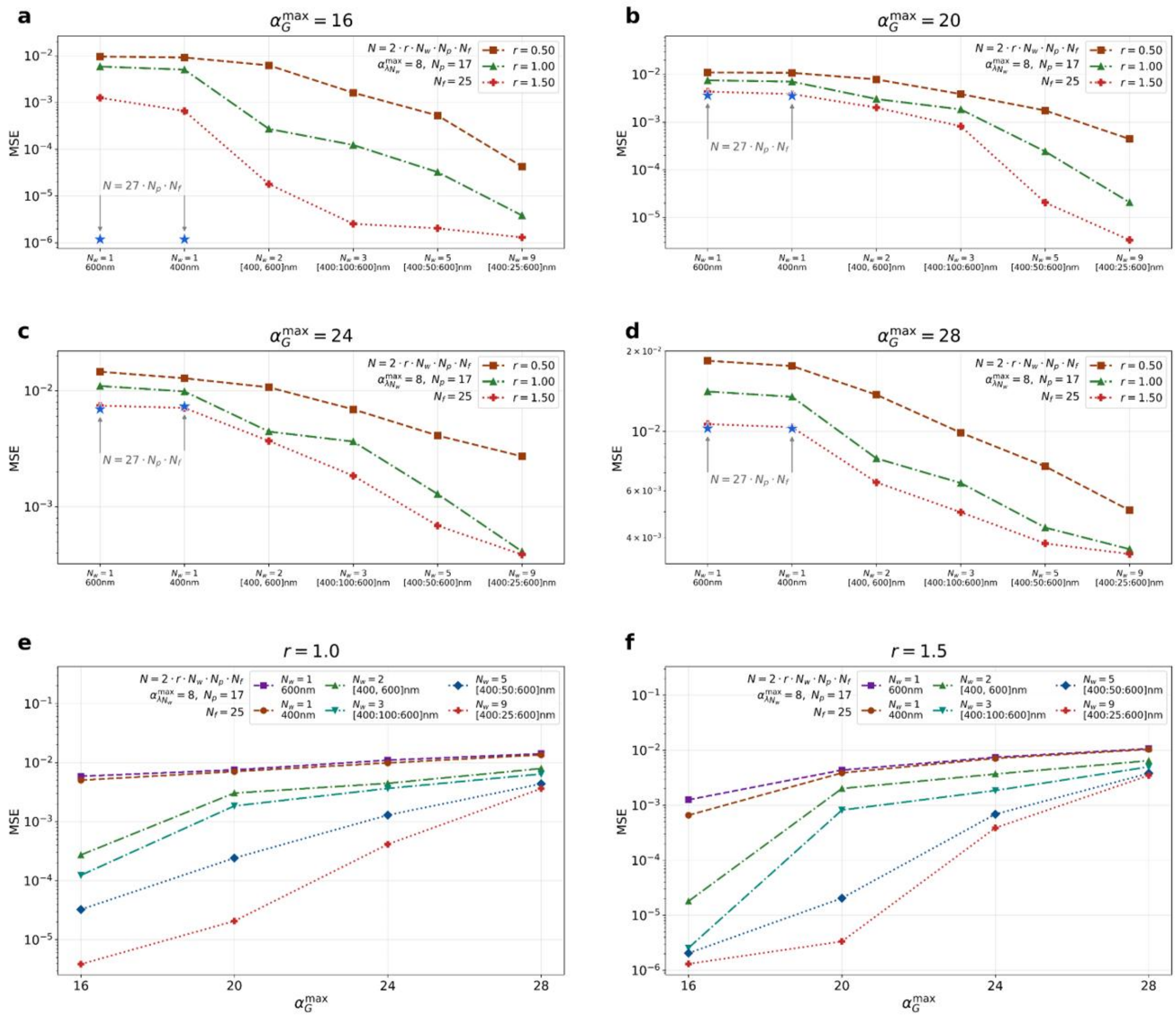

**Fig. 2 Effect of illumination-wavelength diversity on nonlinear function approximation accuracy using spectral-domain diffractive nonlinear processors.** (a)–(d) MSE as a function of the number of simultaneously illuminated wavelengths $N_w$ for target-function maximum frequency values of $\alpha_G^{\max}$ = 16, 20, 24 and 28, respectively. Results are shown for $r$ = 0.5, 1.0 and 1.5, where the total number of trainable diffractive features is $N = 2\ r\ N_w N_p N_f$ (except for the monochrome diffractive designs shown with blue stars, which used $N = 27\ N_w N_p N_f = 27\ N_p N_f$). These designs, shown with the blue stars, indicate that, for higher bandwidth nonlinear target functions, the monochrome diffractive systems with $N_w = 1$ cannot achieve the same function approximation accuracy as multi-wavelength processors, even if they are given substantially increased degrees of freedom, $N$. (e-f) MSE as a function of $\alpha_G^{\max}$ for the different spectral illumination configurations at $r$ = 1.0 and 1.5, respectively. All results use $\alpha_{\lambda_{N_w}}^{\max}$ = 8, $N_p$ = 17, $N_f$ = 25 and $L_K = 3$. Also see **Supplementary**

**Fig. S1** for increased values of $\alpha_{\mathrm{G}}^{\max}$ along with increased number of encoder harmonics, i.e., $N_p = 33$ and $\alpha_{\lambda_{N_w}}^{\max} = 16$.

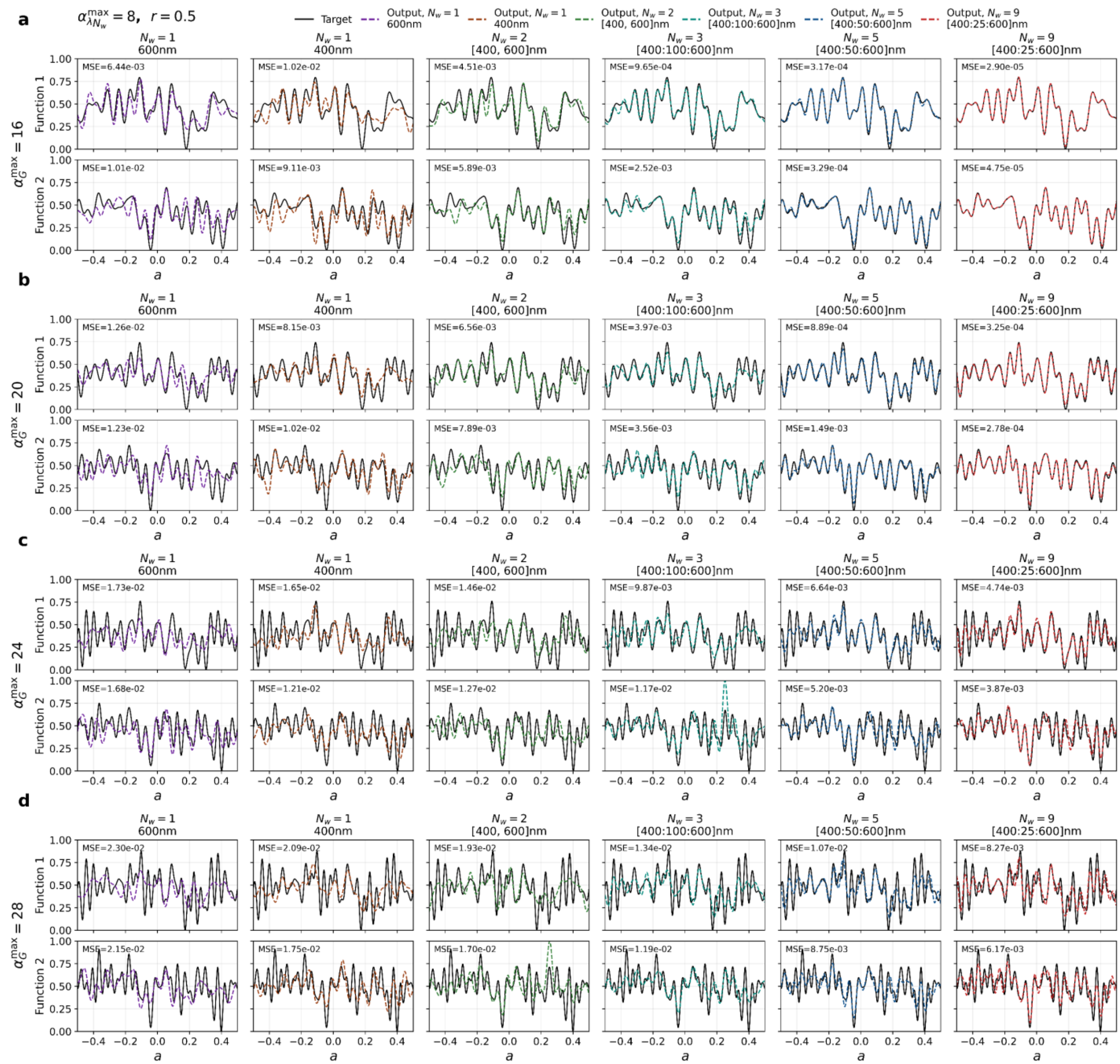

**Fig. 3 Representative nonlinear function approximation results using diffractive nonlinear processors under different spectral illumination configurations; $\alpha_{\lambda_{N_w}}^{\max}$ = 8, $N = N_w N_p N_f$.** (a)–(d) Examples of two target nonlinear functions and their corresponding diffractive-processor outputs for $\alpha_G^{\max}$ = 16, 20, 24 and 28, respectively. The columns compare single-wavelength ($N_w = 1$) illumination at 600 nm and 400 nm with multi-wavelength illumination using $N_w$ = 2, 3, 5 and 9 wavelengths spanning 400–600 nm. In

each panel, the black solid curves denote the target functions and the colored dashed curves denote the corresponding spectral-domain diffractive outputs. All results use $\alpha_{\lambda_{N_w}}^{\max} = 8$, $N_p = 17$, $N_f = 25$, $L_K = 3$ and $N = N_w N_p N_f$, i.e., $r = 0.5$. Also see **Supplementary Figs. S2-S6** for different $\alpha_{\lambda_{N_w}}^{\max}$, $N_p$ and $N$ configurations.

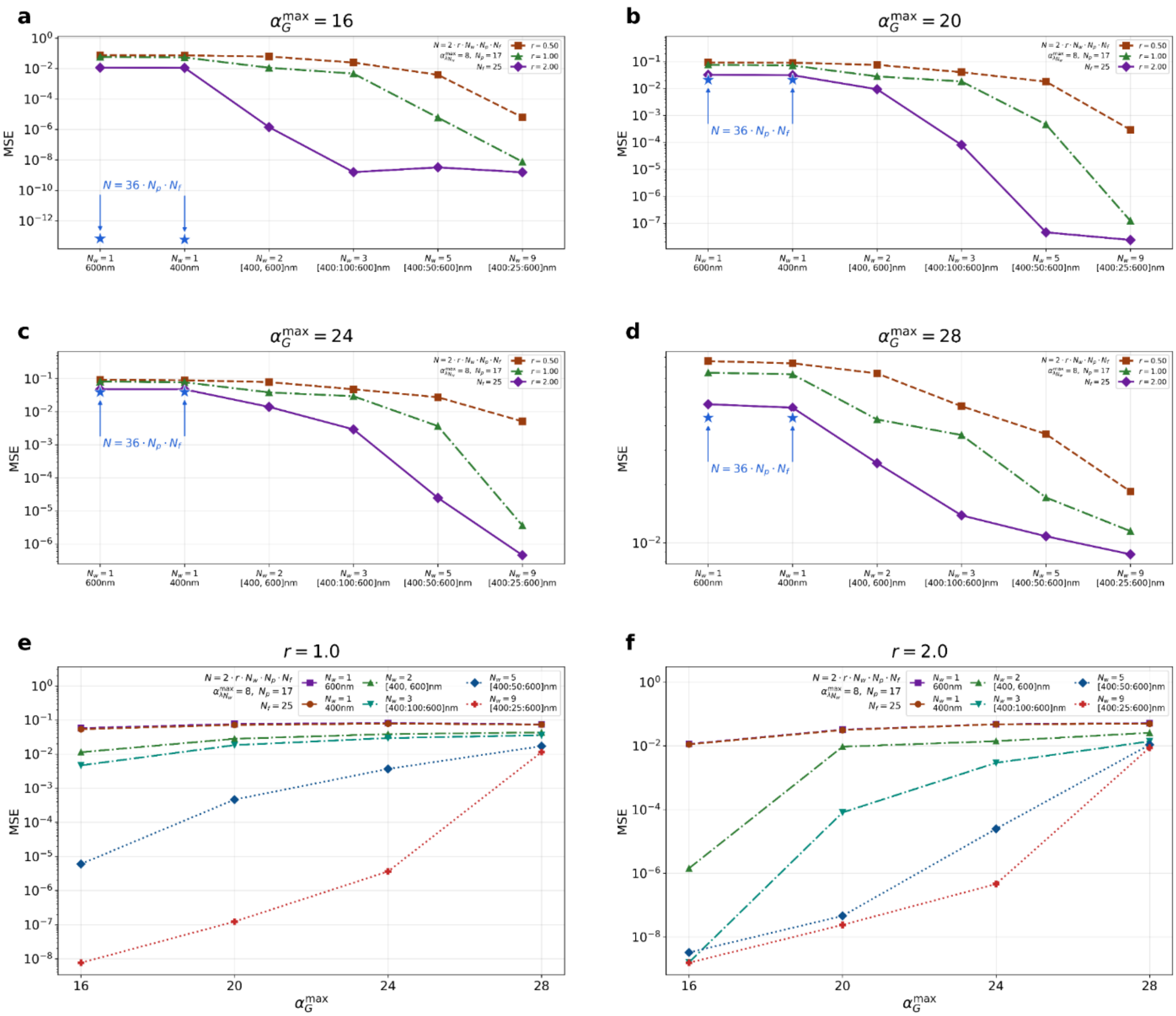

**Fig. 4 Effect of illumination-wavelength diversity on complex-valued nonlinear function approximation accuracy using spectral-domain diffractive nonlinear processors.** Here, three separated detectors approximate distinct real-valued non-negative nonlinear functions, which are combined with bases $e_c = exp(j2\pi c/3)$ to reconstruct the target complex-valued nonlinear function. (a)–(d) MSE as a function of the number of simultaneously illuminated wavelengths $N_w$ for target-function maximum frequency values

of $\alpha_G^{\max}$ = 16, 20, 24 and 28, respectively. Results are shown for $r$ = 0.5, 1.0 and 2.0, where the total number of trainable diffractive features is $N = 2\, r\, N_w N_p N_f$ (except for the monochrome diffractive designs shown with blue stars, which used $N = 36\, N_w N_p N_f = 36\, N_p N_f$). These designs shown with the blue stars indicate that, for higher bandwidth nonlinear target functions, the monochrome diffractive systems with $N_w = 1$ cannot achieve the same function approximation accuracy as multi-wavelength processors, even if they are given substantially increased degrees of freedom, $N$. (e-f) MSE as a function of $\alpha_G^{\max}$ for the different spectral illumination configurations at $r$ = 1.0 and 2.0, respectively. All results use $\alpha_{\lambda_{N_w}}^{\max}$ = 8, $N_p$ = 17, $N_f$ = 25 and $L_K = 3$.

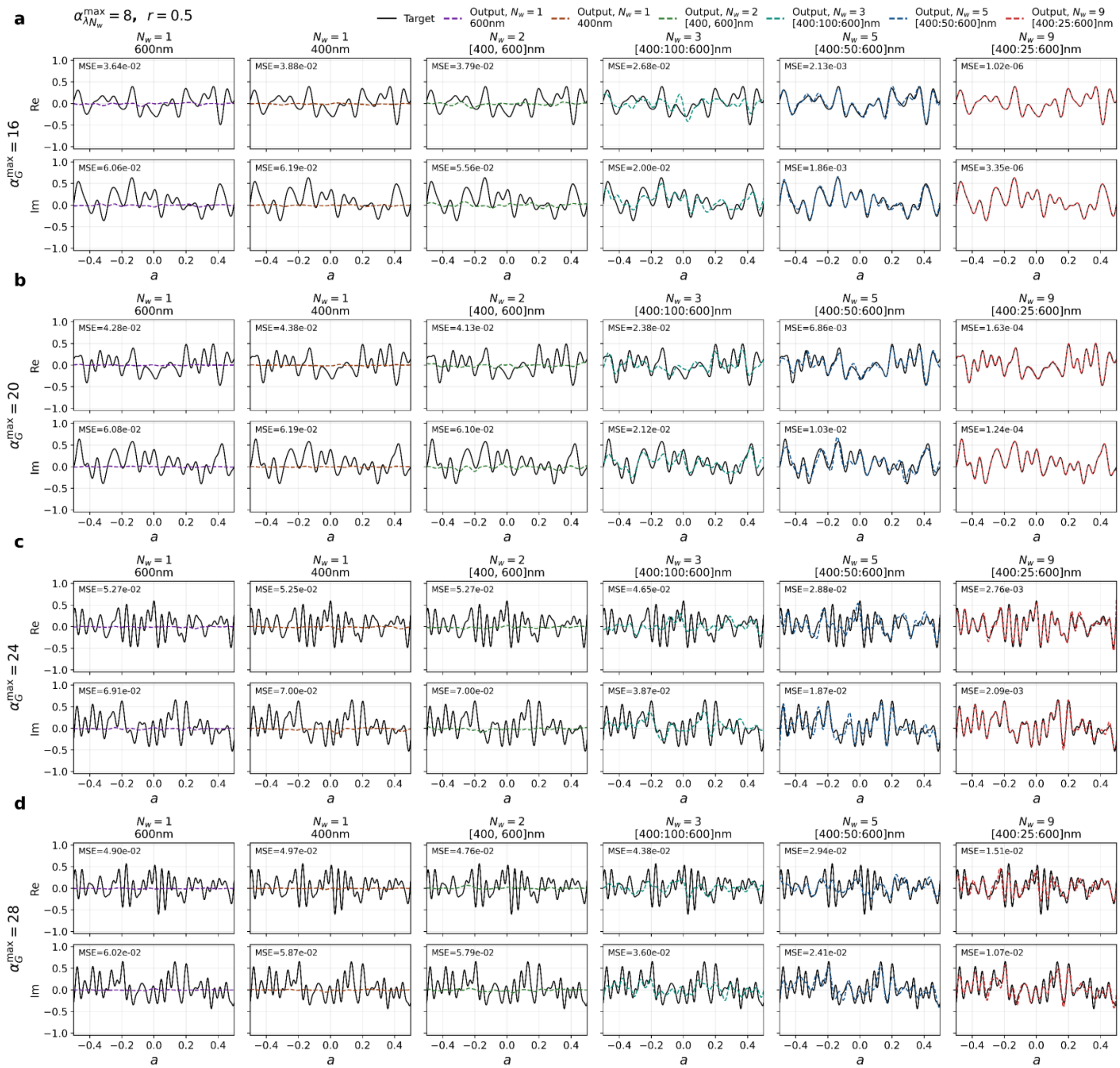

**Fig. 5 Representative complex-valued nonlinear function approximation results using diffractive nonlinear processors under different spectral illumination configurations; $\alpha_{\lambda_{N_w}}^{\max}$ = 8, $N = N_w N_p N_f$.** (a)–(d) Examples of the real and imaginary parts of the target nonlinear functions and their corresponding diffractive-processor outputs for $\alpha_G^{\max}$ = 16, 20, 24 and 28, respectively. The columns compare single-wavelength ($N_w = 1$) illumination at 600 nm and 400 nm with multi-wavelength illumination using $N_w$ = 2, 3, 5 and 9 wavelengths spanning 400–600 nm. In each panel, the black solid curves denote the target functions and

the colored dashed curves denote the corresponding diffractive outputs. All results use $\alpha_{\lambda_{N_w}}^{\max}$ = 8, $N_p$ = 17, $N_f$ = 25, $L_K = 3$ and $N = N_w N_p N_f$, i.e., $r$ =0.5.

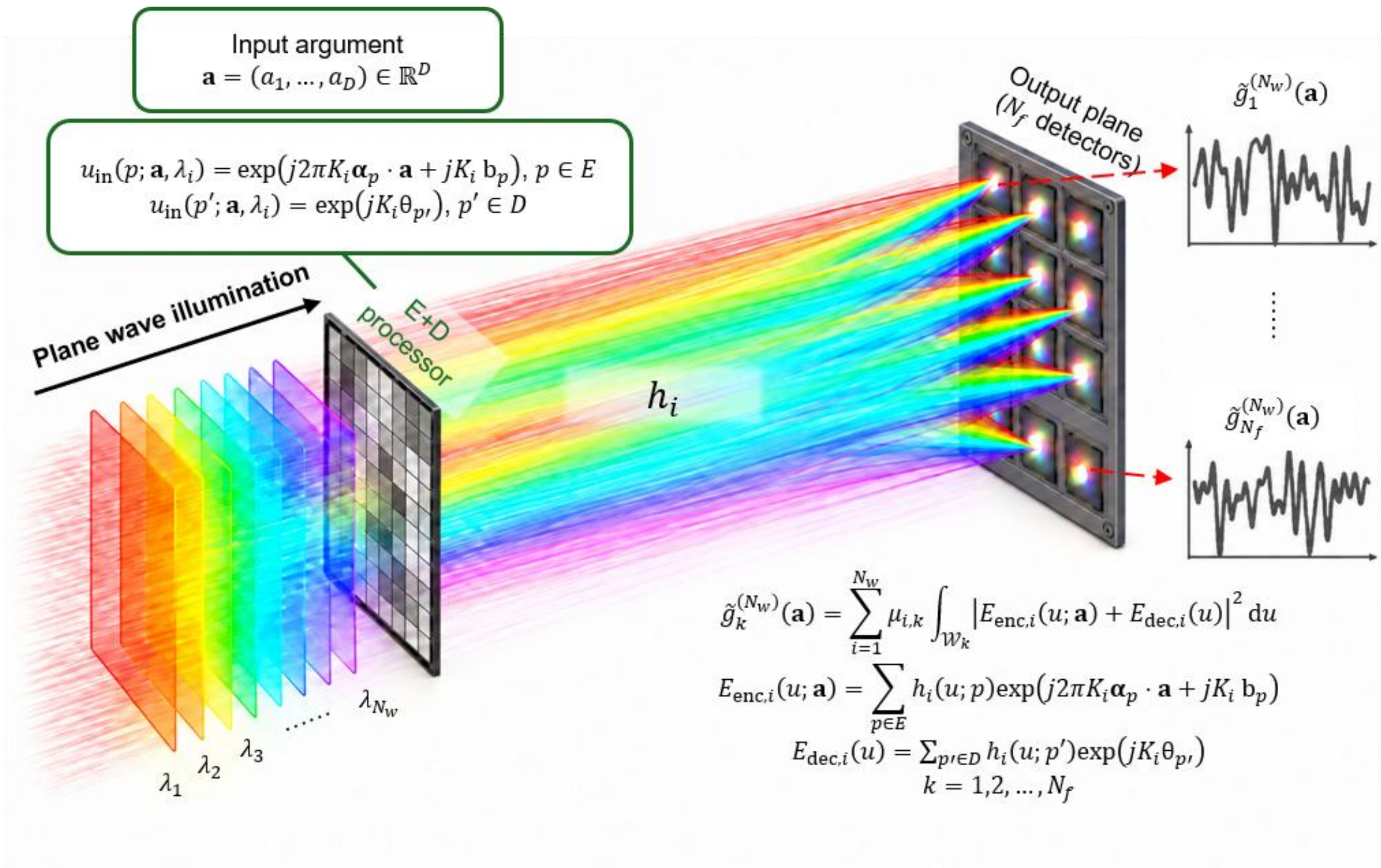

**Fig. 6 Spectral-domain Encoder-Decoder (E+D) co-localized diffractive nonlinear processor.** A set of spatially coherent plane waves at discrete wavelengths $\{\lambda_i\}_{i=1}^{N_w}$ simultaneously illuminates a phase-only E+D input plane comprising an input-dependent encoder and a shared trainable decoder containing $N$ trainable phase features. For an input argument $\mathbf{a} = (a_1, \dots, a_D) \in \mathbb{R}^D$, each encoder feature $p$ applies an argument-dependent phase modulation defined by its computational harmonic, $2\pi K_i \boldsymbol{\alpha}_p \cdot \mathbf{a} + K_i b_p$, whereas each decoder feature $p'$ applies an argument-*independent* trainable phase modulation $K_i \theta_{p'}$. Here, $K_i$ accounts for the wavelength-dependent scaling of the phase modulation at illumination wavelength $\lambda_i$. The optical fields originating from the encoder and decoder channels co-propagate onto an output plane containing $N_f$ detector regions, where they coherently interfere. At wavelength $\lambda_i$, the resulting encoder and decoder fields are denoted by $E_{\text{enc},i}(u; \mathbf{a})$ and $E_{\text{dec},i}(u)$, respectively, and are formed through the wavelength-dependent coherent point-spread functions $h_i(u; p)$ of the E+D processor. For each output channel $k$, the interference intensity is spatially integrated over the corresponding detector region, $W_k$, and the detected intensities from the different wavelengths are combined using wavelength- and detector-dependent scaling factors $\mu_{i,k}$, yielding the nonlinear function output $\tilde{g}_k^{(N_w)}(\mathbf{a}) = \sum_{i=1}^{N_w} \mu_{i,k} \int_{W_k} |E_{\text{enc},i}(u; \mathbf{a}) + E_{\text{dec},i}(u)|^2 \, du$.

An initial version of this illustration was generated using ChatGPT with the authors' text prompt and was subsequently modified by the authors.

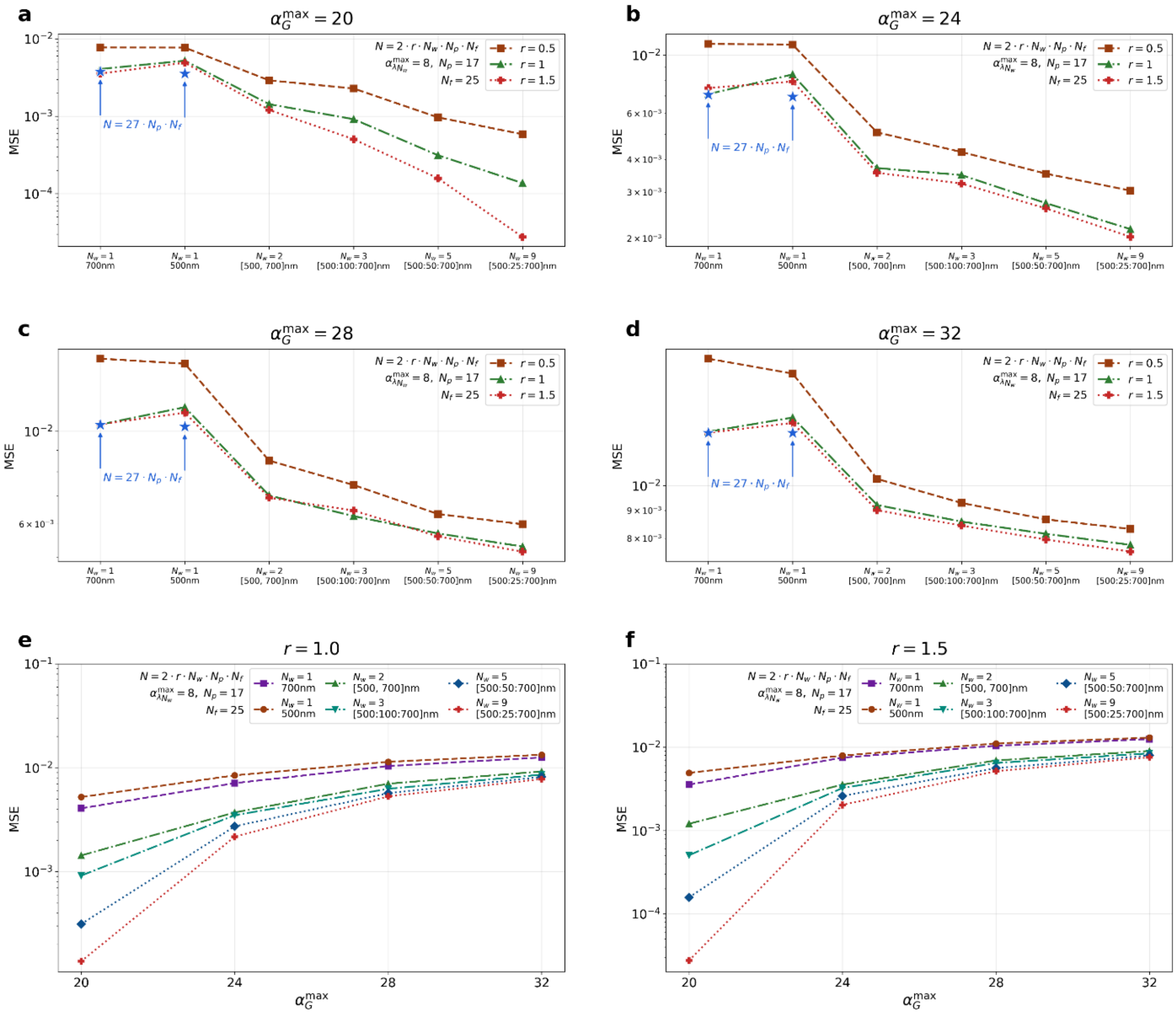

**Fig. 7 Effect of illumination-wavelength diversity on nonlinear function approximation accuracy using spectral-domain E+D diffractive processors.** (a)–(d) MSE as a function of the number of simultaneously illuminated wavelengths $N_w$ for target-function maximum frequency values of $\alpha_G^{\max}$ = 20, 24, 28 and 32, respectively. Results are shown for $r$ = 0.5, 1.0 and 1.5, where the total number of trainable diffractive features is $N = 2\,r\,N_w N_p N_f$ (except for the monochrome diffractive designs shown with blue stars, which used $N = 27\,N_w N_p N_f = 27\,N_p N_f$). These designs shown with the blue stars indicate that, for higher bandwidth nonlinear target functions, the monochrome diffractive systems with $N_w = 1$ cannot achieve the same function approximation accuracy as multi-wavelength

E+D diffractive processors, even if they are given substantially increased degrees of freedom, $N$. (e-f) MSE as a function of $\alpha_G^{\max}$ for the different spectral illumination configurations at $r$ = 1.0 and 1.5, respectively. All results use $\alpha_{\lambda_{N_w}}^{\max}$ = 8, $N_p$ = 17, and $N_f$ = 25.

a

b

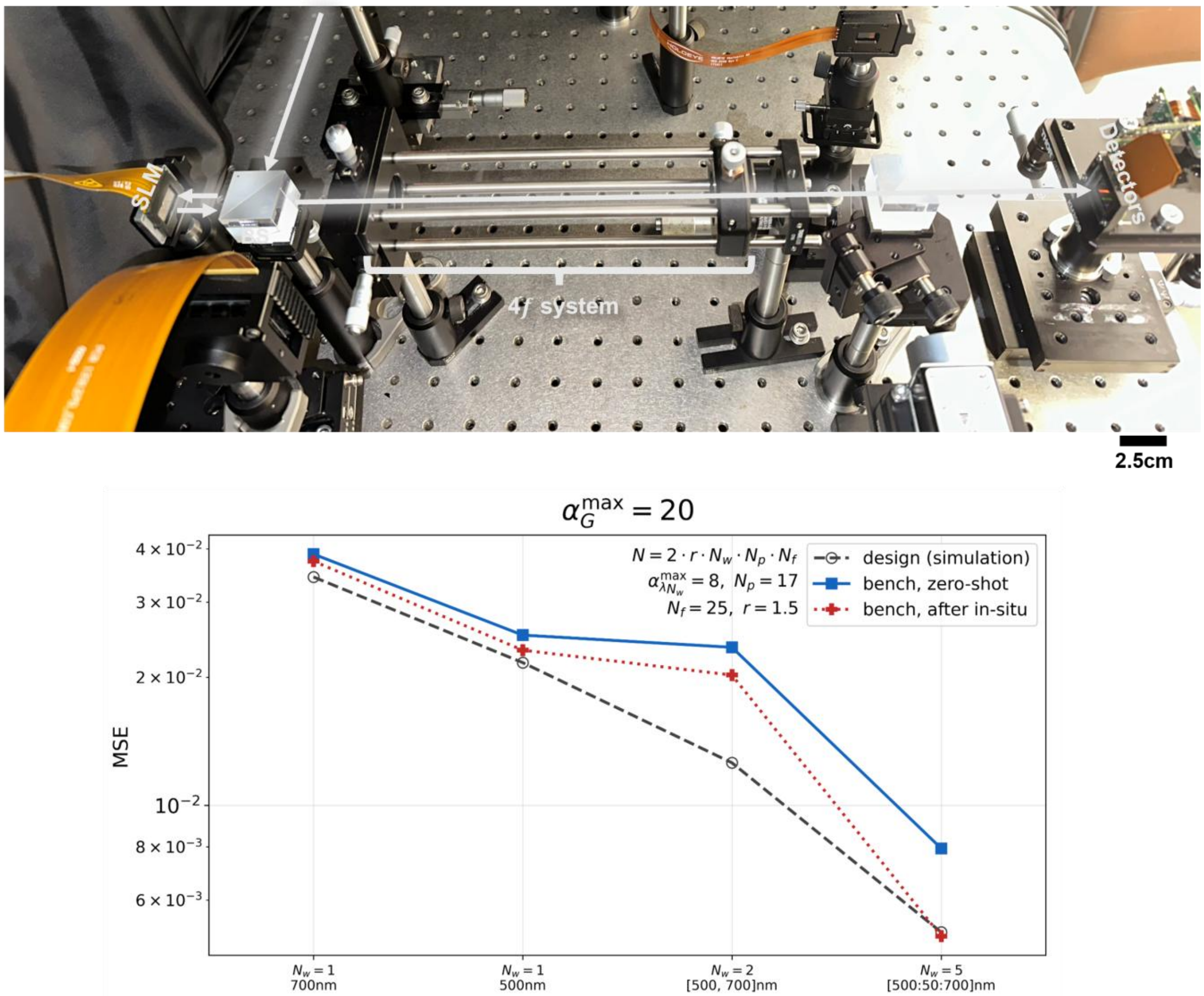

**Fig. 8 Experimental setup and validation of spectral diversity in a spectral-domain E+D diffractive nonlinear processor.** (a) Photograph of the experimental setup, showing the SLM, the 4f relay system, and the output sensor. White arrows indicate the optical path. Scale bar, 2.5 cm. (b) MSE as a function of the number of simultaneously illuminated wavelengths $N_w$ for $\alpha_G^{\max}$ = 20. The total number of trainable diffractive features is $N = 2\, r\, N_w N_p N_f$, $r$ = 1.5. All results use $\alpha_{\lambda_{N_w}}^{\max}$ = 8, $N_p$ = 17, and $N_f$ = 25. Black, blue and red lines show the numerical designs, zero-shot experimental measurements, and experimental measurements after in-situ learning, respectively. For all illumination configurations, the

designs include a penalty on the wavelength-averaged diffraction efficiency falling below a common target of 2.6% (see **Methods** section for the details).

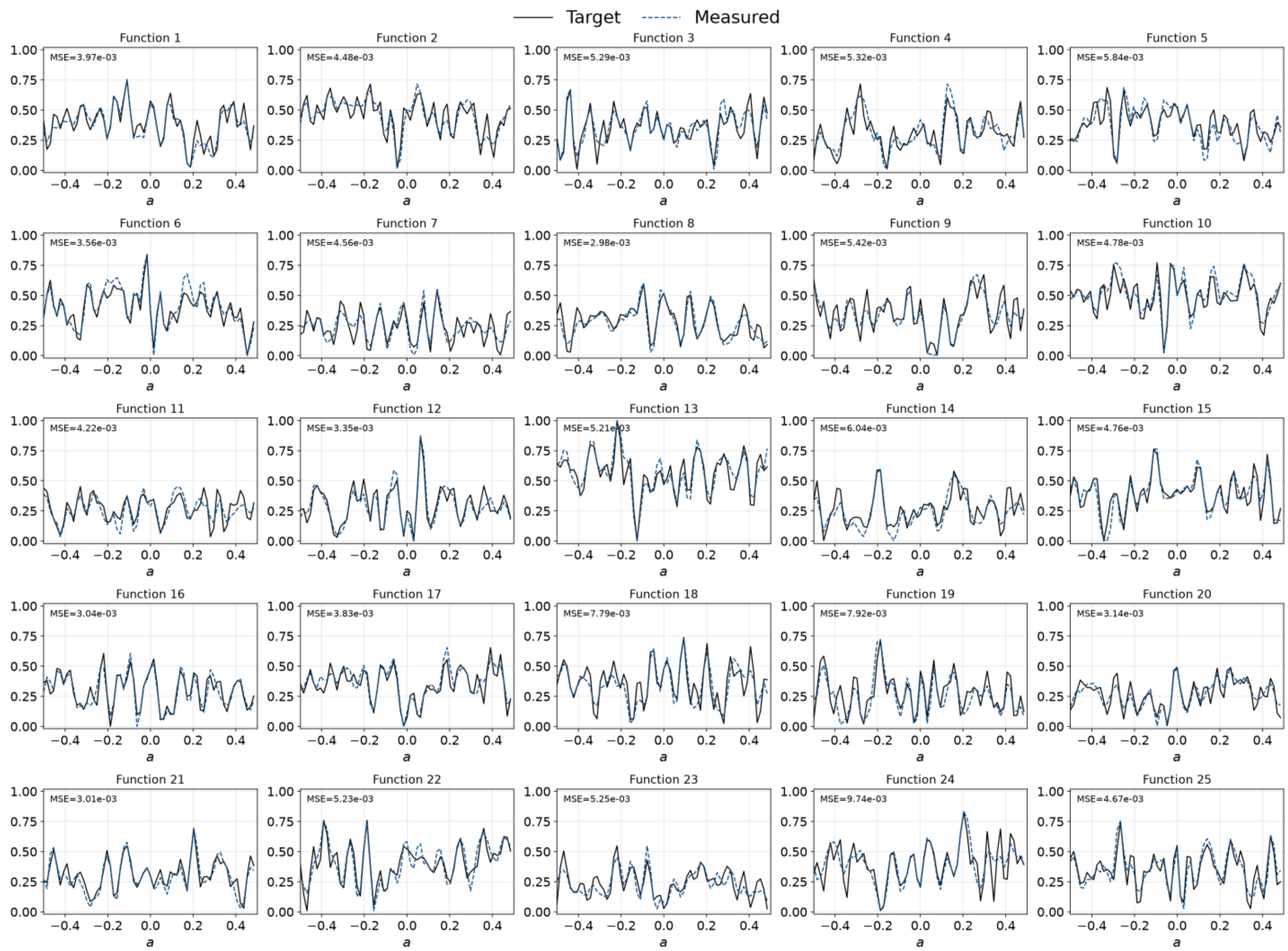

**Fig. 9 Experimental results on nonlinear function approximation using a spectral-domain E+D diffractive processor.** We used $\alpha_{\lambda_{N_w}}^{\max} = 8$, $N_p = 17$, $N_f = 25$, $N_w = 5$, $N = 2rN_wN_pN_f$ and $r = 1.5$. The panels show all $N_f = 25$ target nonlinear functions and their corresponding spectral-domain diffractive processor outputs, implemented in parallel, for $\alpha_G^{\max} = 20$. The blue dashed lines refer to the measurements after in-situ training. The black lines refer to the target nonlinear functions.

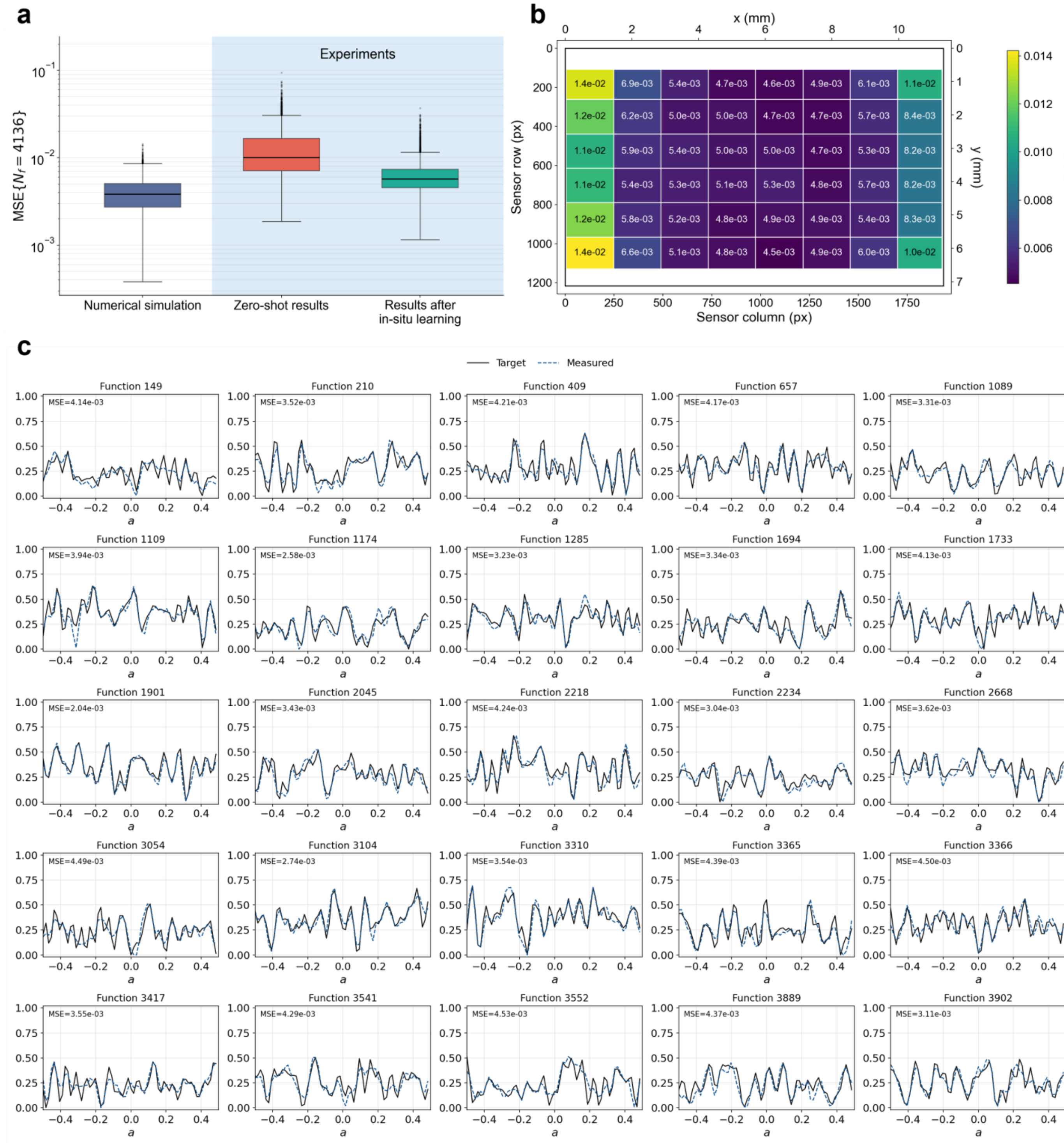

**Fig. 10 Experimental parallel implementation of $N_f = 4{,}136$ nonlinear functions using a spectral-domain E+D diffractive processor.** (a) Distributions of the MSE across all 4,136 target functions for numerical simulations, zero-shot experimental measurements, and experimental measurements after in-situ learning. (b) Spatial distribution of the mean MSE after in-situ learning across detector subregions. Each cell reports the MSE averaged over the function outputs assigned to that subregion, with color indicating the error magnitude. (c) Representative approximation results for target nonlinear functions randomly picked out of $N_f = 4{,}136$ functions, with a maximum target-function frequency of $\alpha_G^{\max} = 20$. Black solid curves denote the target functions, and blue dashed curves denote the measured outputs after in-situ training. The processor uses $\alpha_{\lambda_{N_w}}^{\max} = 8$, $N_p = 17$, $N_w = 5$, and $N_f = 4{,}136$. Also see **Supplementary Figs. S13-S15** for the additional examples of the $N_f = 4{,}136$ nonlinear functions.